\documentclass[fleqn,10pt]{my_wlscirep}

\usepackage{graphicx}
\usepackage{dcolumn}
\usepackage{bm}

\usepackage{amsmath,amssymb,psfrag,graphicx}

\usepackage{multibib}
\usepackage{multirow}
\usepackage[pagewise]{lineno}
\modulolinenumbers[2]

\newcites{tex}{References}
\newcites{SI}{Additional references}

\def\ds{\displaystyle}
\def\ss{\scriptstyle}

\def\eps{\varepsilon}

\newcommand{\onehalf}{\mbox{$\frac{\ss 1}{\ss 2}$}}

\newcommand{\onefourth}{\mbox{$\frac{\ss 1}{\ss 4}$}}

\newcommand{\Frac}[2]{\ds\frac{#1}{#2}}
\title{Rethinking network reciprocity over social ties:\\
local interactions enable direct reciprocity and pave the rational way to cooperation}

\author[1,*]{Fabio Dercole}
\author[1]{Fabio Della Rossa}
\author[1]{Carlo Piccardi}
\affil[1]{Department of Electronics, Information, and Bioengineering, Politecnico di Milano, Piazza Leonardo da Vinci 32, I-20133, Milano, Italy}
\affil[*]{fabio.dercole@polimi.it}

\newcommand{\onlinecite}[1]{\hspace{-1 ex} \nocite{#1}\citenum{#1}} 

\makeatletter
\newcommand{\customlabel}[2]{%
	\protected@write \@auxout {}{\string \newlabel {#1}{{#2}{}{}{equation.M1}{}}}}
\makeatother


\begin{abstract}

\vspace{-7mm}\noindent
Since Nowak \& May's (1992) influential paper, {\it network reciprocity}---the fact that local interactions allow unconditional cooperators to self-organize into clusters---was proposed as the simplest mechanism for the evolution of cooperation in biological and socio-economic systems.
It has been confirmed in several theoretical models and shown to predict the fixation of cooperation under a simple rule:
the benefit of a single altruistic act must outweigh the cost of cooperating with all neighbors.
The experimental evidence among humans is however controversial.
The reason is that, while models assume that individuals update strategy by imitating better performing neighbors, experiments showed that humans are more prone to reciprocate cooperation than to compare payoffs.
Motivated by the empirical results, we rethink network reciprocity as a rational form of direct reciprocity on networks, indeed made possible by the locality of interactions.
Imitating a better performing neighbor is irrational in networks far from all-to-all, as it can well cause a payoff loss.
Instead, we base strategy update on a model prediction among two options:
conditional---reciprocal---cooperation and unconditional defection in a networked prisoner’s dilemma.
We show that both reciprocity and a multi-step predictive horizon are necessary to sustain cooperation, as well as sufficient for its fixation in any network, provided the game benefit-to-cost ratio is larger than a measure of network’s connectivity.
We hence rediscover the same simple rule---experimentally validated---underpinned however with a different evolutionary mechanism.
\end{abstract}

\begin{document}
	
\rfoot{\small\sffamily\bfseries\thepage/\pageref{finalpagearticle}}%
\flushbottom
\maketitle


\vspace{-9mm}
\section*{Introduction}
Cooperation among rational---self-interested---agents is a longstanding and still debated puzzle in biology and social sciences, with countless contributions since Axelrod, Hamilton, and Trivers' seminal works, \citetex{Trivers_71_QRB,Axelrod81}
and the topic received recent attention also in several fields of engineering.
\citetex{IEEE_CS_special_GT}

The standard modeling framework is {\it evolutionary game theory} (EGT), \citetex{Hofbauer_and_Sigmund_98_CUP} in which a non-cooperative game---where any altruistic act is self-enforcing---describes a repeated interaction among pairs (or larger groups) of individuals, a given set of strategies is confronted, and an evolutionary process links the obtained payoffs to reproduction and death in biology or to strategy-update in socio-economic systems.
The paradigmatic game used to study the evolution of cooperation is the {\it prisoner's dilemma} (PD)---the two-player-two-option interaction in which a cooperator (option C) provides a benefit $b$ to the opponent at a cost $c\hspace{-0.5mm}<\hspace{-0.5mm}b$ to herself, whereas a defector (option D) provides no benefit at no cost.
The {\it benefit-to-cost} ratio, or {\it game return} $r=b/c$, is often used to parameterize the game (taking $c=1$ as monetary unit).

To test whether a cooperative strategy (strategy C) has any chance to evolve, it is confronted with the benchmark strategy `unconditional defection' (strategy D, played by individuals who always defect) under one or a few evolutionary processes.
The three different issues to be discussed are the {\it invasion} of the strategy---the spreading of cooperators (C-strategists) in a population dominated by defectors (D-strategists)---its {\it persistence}---the long-term presence, fluctuating or not, of C's---and its {\it fixation}---the convergence to the state all-C.
For example, it is well known that when the PD is played in large and well-mixed---{\it unstructured}---populations, there is no hope for the strategy `unconditional cooperation' (played by individuals who always cooperate).
Defecting gives the largest payoff regardless of what the others are doing, so that, without any specific incentive to cooperation, C's cannot invade under any reasonable evolutionary process and disappear if initially present in the population.
Compared to other social dilemmas, the PD is the worst-case scenario for the evolution of cooperation.

Traditional incentivizing mechanisms \citetex{Nowak_06_SCI} either make cooperation conditional---such as
{\it reciprocal altruism} \citetex{Trivers_71_QRB,Axelrod81} (also known as {\it direct reciprocity}),
the establishment of {\it reputations} \citetex{Nowak_and_Sigmund_98_NAT} (also known as {\it indirect reciprocity}), and
mechanisms of {\it kin} \citetex{Hamilton64a} or {\it group selection} \citetex{Wilson75b}
(or other forms of {\it assortment} \citetex{Jansen_and_vanBaalen_06_NAT,Fletcher_and_Doebeli_09_PRSB})---or change the rules of the game, as by introducing {\it volunteering} (optional participation) \citetex{Hauert_et_al_02_SCI}
and {\it punishment} of antisocial behaviors \citetex{Boyd_et_al_03_PNAS,Dercole_et_al_13_JTB}.
All these mechanisms add degrees of strategical complexity, either in terms of players' cognitive abilities and/or information flows, or due to extra options in the underlying game.

\begin{figure}[t!]
\centering
\includegraphics{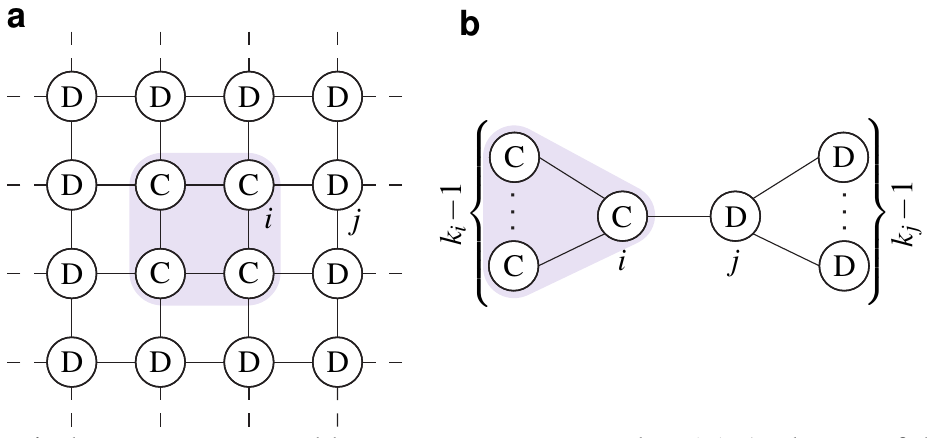}
\caption{Network reciprocity in homogeneous and heterogeneous networks.
({\sffamily\textbf a})
A cluster of $4$ C's surrounded by D's in a square lattice with periodic boundary conditions.
The payoff (per game round, obtained by each individual by playing a PD with all neighbors and summing up outcomes) of C's is $2r-4$;
that of the D's at the border of the cluster is $r$.
C's do better than D's if $r>4$ and the payoff of a D copying a C (e.g., $j$ copying $i$) drops to $r-4$.
({\sffamily\textbf b})
A cluster of $k_i$ C's protected from $k_j$ D's
($k_i,k_j>2$; $k_i$ and $k_j$ are the degrees of nodes $i$ and $j$).
The payoff of the C-individual $i$ is $\pi_i=r(k_i-1)-k_i$; that of the D-individual $j$ is $\pi_j=r$;
$\pi_i>\pi_j$ if $r>1+2/(k_i\hspace{-0.2mm}-\hspace{-0.2mm}2)$ and $j$ drastically reduces her payoff when copying $i$.
Under imitation update and $\pi_i>\pi_j$, $j$ alternates between C (when selecting $i$ for comparison) and D (when selecting a D-neighbor); the other nodes do not change strategy.}
\vspace{-3mm}
\label{fig:net}
\end{figure}

Starting with Nowak and coauthors' influential papers \citetex{Nowak_and_May_92_NAT,Lieberman_et_al_05_NAT,Ohtsuki_et_al_06_NAT}, the fact that interactions in real populations are {\it structured} according to the individuals' personal contacts has been proposed as the simplest mechanism---requiring no strategical complexity---to explain cooperation.
It has been named {\it network reciprocity}, because in a static (SI note 1) and sparse (far from complete) network, unconditional C's can do better than unconditional D's by grouping into clusters, i.e., the network's structure allows C's to reciprocate (Fig.~\ref{fig:net}).
More specifically, in large regular networks---each node having {\it degree} $k$; $k$ neighbors---driven by an imitation-like evolutionary process---individuals imitate from time-to-time a better performing neighbor (SI note 2)---and in the limit of weak selection---the game payoff marginally impacting the individual performance (SI note 3)---unconditional C's can invade and fixate under a simple condition \citetex{Ohtsuki_et_al_06_NAT}: the game return $r$ must exceed the connectivity $k$.
That is, if $r>k$ in the above setting, the probability that cooperation invades and fixates starting from a single C placed in a random position---the fixation probability---is larger than $1/N$---the fixation probability under a totally random process of strategy update, $N$ being the network's size.
And the rule has been generalized to non-regular networks, \citetex{Konno_11_JTB,Allen_et_al_17_NAT} essentially requiring $r$ to exceed the average degree $\langle k\rangle$.

As several authors did in the last decade,
\citetex{Grujic_et_al_10_PONE,gracia2012heterogeneous,gracia2012human,Semmann_12_PNAS,Grujic_et_al_14_SR}
we question network reciprocity as a mechanism supporting cooperation in socio-economic networks.
In particular, we question the rationale behind the imitation process of strategy update.
Why should we copy a better performing neighbor whose neighborhood might be considerably different---in size as well as in composition---from ours?
Especially in heterogeneous networks, imitation may turn counterproductive (see Fig.~\ref{fig:net}, where individual $j$ reduces her payoff by copying $i$).
Recent experiments on (impressively large) human networks playing a PD \citetex{Grujic_et_al_10_PONE,gracia2012heterogeneous} indeed showed that we are more prone to reciprocate cooperation than to compare payoffs.
Specifically, the probability to cooperate
(as a rule of the experiment, the same game option, C or D, is taken toward all neighbors)
is conditioned by the player's previous choice, i.e., by the player's cooperative or defective `mood.' \citetex{gracia2012human,Grujic_et_al_14_SR}
In the C mood, a player is willing to cooperate the more cooperation she observed in the previous game round, even though the neighbors' previous payoffs were made available;
in the D mood, defection is rather unconditional (or weakly correlated with the previous level of cooperation).
Nonetheless, the simple rule of network reciprocity, $r>\langle k\rangle$, found empirical validation. \citetex{Rand_et_al_14_PNAS,Li_et_al_18_PNAS}

Motivated by the empirical findings, we rethink the role of the population structure.
In the C mood, players showed a form of direct reciprocity, \citetex{Trivers_71_QRB,Axelrod81} i.e., higher chance to cooperate with neighbors who showed altruism in past interactions.
As all forms of reciprocity, it requires repeated interactions with the same individuals as well as cognitive abilities to recognize individuals and remember past interactions.
This is exactly the environment provided by a static and sparse network.
In particular, it is the locality of interactions that makes cognitive tasks possible, as the required resources---in terms of memory and abilities---scale with the player's neighborhood.
At the same time, a local interaction opens the way to more complex rules of strategy update, not only based on past interactions, but also on foreseeing future ones.
We theoretically test reciprocal cooperation against the benchmark `unconditional defection' under a rational process of strategy update, based on a model prediction of their future income.
Reciprocity is implemented by allowing C's to selectively abstain from playing for a few game rounds with exploiting neighbors.
\citetex{Szabo_and_Hauert_02_PRE,Hauert_et_al_02_SCI,Li_et_al_17_SR}
So doing, C's reduce exploitation risks---with respect to unconditional C's---and, at the same time, communicate their mood, thus increasing their chances to reciprocate cooperation in future interactions.
We name this new mechanism for the social evolution of cooperation {\it networked rational reciprocity}.

Our main result is that, under a rule qualitatively similar to $r>\langle k\rangle$, networked rational reciprocity grants the fixation of cooperation starting from any cluster of two C's.
The more interactions are local (the sparser is the network), the lower is the required return $r$.
And even starting from a single C the fixation probability remains high the more connected is the initial C, highlighting the role of the network's structure.
We hence rediscover the simple rule of network reciprocity, but we provide a different underlying explanation, more in line with the observed social behavior.





\section*{Results}

\subsection*{Model description}
\label{sec:mod}
Before presenting results, we introduce our model
(implementation details are given in the Methods section;
the elements of novelty are commented in the Discussion).
We consider two strategies, a reciprocating form of conditional cooperation (strategy C) and unconditional defection (strategy D).
At each game round, each individual is a C- or a D-strategist, the strategy representing the individual's mood.
A PD is played by all pairs of connected individuals.
The individual payoff in the round is the sum of the outcomes of the PD interactions in the neighborhood.
After each round, each individual revises her strategy with a probability $\delta$ assumed small and uniform across the population;
parameter $\delta$ measures the rate (per game round) of (asynchronous) strategy update, and $1/\delta$ is the average number of rounds between two consecutive revisions by the same individual, a sort of `inertia' to change.

In each PD interaction, a C can selectively opt for cooperation or abstention;
specifically, she stops playing with an exploiting neighbor for a number of rounds drawn in accordance with the probability that the exploiter revises her strategy just after.
D's always play and defect.
Our conditional C-strategy implements a form of direct reciprocity.
As in the famous tit-for-tat strategy, C's cooperate with neighbors who showed cooperation in the previous round.
However, they do not retaliate for defection; rather, they abstain from playing, communicating their mood to the opponent;
moreover, they forgive defection or, better, they poll previous exploiters to seek for cooperation.
To modulate reciprocity in the population, the length of abstention periods is drawn according to a reciprocity-biased rate of strategy update\\[-2mm]
\begin{equation}
\label{eq:deps}
\delta_{\eps}=(1-\eps)\delta\in(0,1),
\end{equation}
where parameter $\eps$, also assumed small and uniform across the population, measures reciprocity;
super/sub reciprocating C's ($\eps\gtrless 0$) wait longer/shorter, on average w.r.t. normally reciprocating ones ($\eps=0$), to go back playing.

When revising strategy, an individual computes her expected accumulated payoff behaving as C or D during an horizon of $h$ future interactions and decides for the more profitable strategy until the next update.
The prediction is based on the model society here described that is assumed to be public knowledge.
We consider the minimal-information scenario, in which individuals have no access to neighbors' payoffs and connectivity and infer the neighbors' strategies only from past interactions.
Consequently, the model prediction cannot account for the concomitant changes in the neighbors' strategies.
This limits the horizon to a few rounds under a relatively slow strategy update (small $\delta$).
Because of the short horizon, no discount of future incomes is adopted.

The model parameters and their numerical values used in the analysis are summarized in Table~\ref{tab:par}.

\subsection*{Analytical results}
\label{sec:ares}
To gain insight in the system's dynamics, we preliminary consider an infinite predictive horizon, because it allows a simpler analysis.
We prove (in SI
Sects.~\ref{sec:pt}--\ref{sec:hinf}) that when a C-player with degree $k$ and $k_{\mathrm{C}}$ C-neighbors (known from past interactions) revises her strategy according to an infinite horizon, she remains C (no expected gain in changing to D) if
\begin{equation}
\label{eq:condrinf}
r>1+\Frac{k}{k_{\mathrm{C}}}\Frac{P_{\mathrm{CD}}^{\infty}}{1-P_{\mathrm{CD}}^{\infty}},\quad
P_{\mathrm{CD}}^{\infty}=\Frac{1}{2}\Frac{\sqrt{4\delta_{\eps}-3\delta_{\eps}^2}-\delta_{\eps}}{1-\delta_{\eps}}
\approx\sqrt{\delta_{\eps}}\;\;\text{for small $\delta_{\eps}$},
\end{equation}
where $P_{\mathrm{CD}}^{\infty}$ is the probability (computed in Sect.~\ref{sec:pCDh}) that a C-player who remains C forever will get
exploited by a D-neighbor who remains D forever in a far-future interaction.
Similarly, under the same condition~\eqref{eq:condrinf}, the D-to-C strategy change occurs (positive expected gain in changing to C) according to an infinite predictive horizon.

With a finite horizon of $h\ge 1$ future interactions, the conditions governing strategy update are more complex
(see
Sects.~\ref{sec:DpiC} and~\ref{sec:DpiD}, where the expected gains $\Delta\pi_{\mathrm{C}}^h$ and $\Delta\pi_\mathrm{D}^h$ respectively predicted by a C and a D for a strategy change are computed).
Given a C and a D with identical neighborhoods, the $r$-threshold for C to remain C and that for D to change to C are different.
Typically, the former is lower, as the C-neighbors of a C are more prone to play in the near future
(see
Sect.~\ref{sec:rch}).
Not surprisingly, for $h=1$ (best-response update\citetex{Roca2009}) we have $\Delta\pi_{\mathrm{C}}^1>0$ and $\Delta\pi_\mathrm{D}^1<0$, i.e.,
defecting assures the highest payoff independently of the network's structure.
Moreover, under a condition on $r$ more restrictive than \eqref{eq:condrinf}
(derived in
Sect.~\ref{sec:cond2}), the predicted payoff gains $\Delta\pi_{\mathrm{C}}^h$ and $\Delta\pi_\mathrm{D}^h$ are $h$-monotonic, respectively decreasing and increasing to the negative and positive infinite-horizon limits.
For intermediate $r$, $\Delta\pi_{\mathrm{C}}^h$ (resp.~$\Delta\pi_\mathrm{D}^h$) first increases (decreases) with $h$ up to a positive (negative) extremum, then decreases (increases) to the negative (positive) infinite-horizon limit.
Finally, we show
(Sect.~\ref{sec:cond2}) that the effect on $\Delta\pi_{\mathrm{C}}^h$ (resp.~$\Delta\pi_\mathrm{D}^h$) of adding one prediction step can be made arbitrarily negative (positive) by a sufficiently large $r$.

\begin{figure}[t!]
\centering
\includegraphics{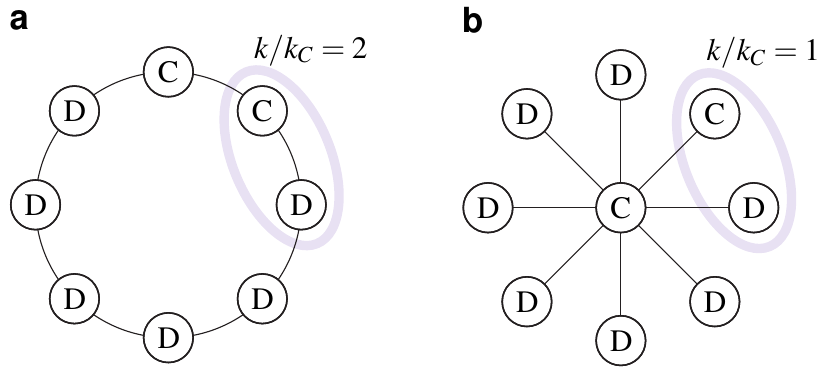}
\caption{Networked rational reciprocity in homogeneous and heterogeneous networks.
Consider ({\sffamily\textbf a}) the ring ($k=2$ for all nodes) and {\sffamily\textbf b} the star network ($\langle k\rangle\approx 2$ for large $N$) in the infinite-horizon limit.
The ring-$r_{\mathrm{C}}^{\infty}$ is given by~\eqref{eq:condrinf} with $k/k_{\mathrm{C}}=2$ and a single D drives the population to all-D if $r<r_{\mathrm{C}}^{\infty}$.
The star-$r_{\mathrm{C}}^{\infty}$ is much higher, because the ratio $k/k_{\mathrm{C}}$ peaks at $N-1$ for the central node with only one C-neighbor.
However, if the central node is a C, the D-leaves will change to C under the weakest requirement on $r$ ($k/k_{\mathrm{C}}=1$ in~\eqref{eq:condrinf}) and there are high chances that this occurs before the central node revises her strategy.
(The probability that a D-leaf revises
before the central node is given by the formula at point (ix), in which $k$ is replaced by the number of D-leaves.)
If the number $k_{\mathrm{C}}$ of C-leaves raises to satisfy condition~\eqref{eq:condrinf}, the population then evolves to all-C, and this occurs with probability higher than $1/2$ starting with no C-leaves and $r$ equal to the ring-$r_{\mathrm{C}}^{\infty}$
(the probability goes as $1/2+1/(2N)$ for small $\delta$, see
Sect.~\ref{sec:pup}).
That is, on average, the isolated central C drives the star to all-C under a milder condition on $r$ w.r.t. the ring with some initial D's.}
\label{fig:ex}
\end{figure}

Despite the system's complexity, the above results have straightforward consequences.
\begin{itemize}
\item[(i)]
The state all-C is invariant under a weak requirement on $r$
(condition~\eqref{eq:condrinf} with $k_{\mathrm{C}}=k$; recall that $\eps$ and $\delta$ are small).
\item[(ii)]
The state all-D is invariant
(condition~\eqref{eq:condrinf} with $k_{\mathrm{C}}=0$),
though a coordinated switch to C by a small cluster of players can give a payoff gain.
\item[(iii)]
Isolated C's ($k_{\mathrm{C}}=0$) change to D as soon as they revise their strategy.
This does not mean that cooperation cannot start from isolated individuals, since D-neighbors could change to C before the isolated C changes
to D.
\item[(iv)]
Indeed, a D player connected to a single C ($k_{\mathrm{C}}=1$) can change to C, provided the game return $r$ is sufficiently large.
\item[(v)]
Direct reciprocity
($\eps>\eps_{\min}=-(1-\delta)/\delta$, so that $\delta_{\eps}<1$)
is essential for the evolution of cooperation, otherwise $P_{\mathrm{CD}}^{\infty}=1$ in~\eqref{eq:condrinf} and D is the well-known best option.
Normal or super reciprocity ($\eps\ge 0$) further helps cooperation.
\item[(vi)]
A multi-step predictive horizon ($h\ge 2$) is essential for cooperation
(see the above discussion on $\Delta\pi_{\mathrm{C}}^1$ and $\Delta\pi_\mathrm{D}^1$).
\item[(vii)]
The inertia to change also helps cooperation, in the sense that a lower rate $\delta$ of strategy update
reflects in longer C's abstentions and hence in a lower $P_{\mathrm{CD}}^{\infty}$ in~\eqref{eq:condrinf}.
\item[(viii)]
Main result: with both reciprocity and multi-step horizon, given all other details,
there is a threshold $r_{\mathrm{C}}^h$ on $r$ above which cooperation fixates starting from any cluster of (at least two) C's.
An upper bound to $r_{\mathrm{C}}^h$ is obtained by considering the most connected node with one C-neighbor
(Sect.~\ref{sec:rch}).
\item[(ix)]
The previous result probabilistically holds also starting from an isolated C, provided her degree $k$ is not too small.
With $r>r_{\mathrm{C}}^h$, the probability that a D-neighbor changes
to C before the isolated C changes
to D goes as $1-1/(k+1)$ for small $\delta$
(Sect.~\ref{sec:pup}).
\item[(x)]
Increasing the horizon $h$ always reduces $r_{\mathrm{C}}^h$
(Sect.~\ref{sec:rch}).
\item[(xi)]
According to condition~\eqref{eq:condrinf}, cooperation seems to be favored in homogeneous sparse networks (low largest degree) compared to heterogeneous and/or dense ones.
This is evident at low levels of cooperation, at which adding links to a node is not likely to increase the number of its C-neighbors, thus increasing the ratio $k/k_{\mathrm{C}}$.
\item[(xii)]
In the complete network---to be used as benchmark since direct reciprocity is there unfeasible---the threshold $r_{\mathrm{C}}^h\!$ is maximal.
%
\item[(xiii)]
The role of network's structure:
%
the comment at  point (ix) and two simple examples in Fig.~\ref{fig:ex} suggest that degree heterogeneity helps cooperation only if C's initially occupy the network's hubs.
Essentially, low-degree D's connected to a C change to C under a mild requirement on $r$ (low ratio $k/k_{\mathrm{C}}$ in~\eqref{eq:condrinf}).
However, to have many low-degree D's connected to few initial C's, we need high-connected C's.
Hence, especially starting at low levels of cooperation, degree heterogeneity and the placement of the initial C's in the network's hubs together reduce the required $r$ to evolve to all-C.
Compared to imitation update (in which C-hubs need a significant fraction of C-neighbors to persist), our rational process of strategy update allows the formation of C-clusters even starting from isolated C-hubs, who pay (or better invest in) the initial cost of exploitation.
If however most of the hubs are D's, network heterogeneity turns harmful to cooperation (see point (xi)).
%
\end{itemize}

\subsection*{Numerical results}
\begin{figure}[t!]
\centering
\includegraphics{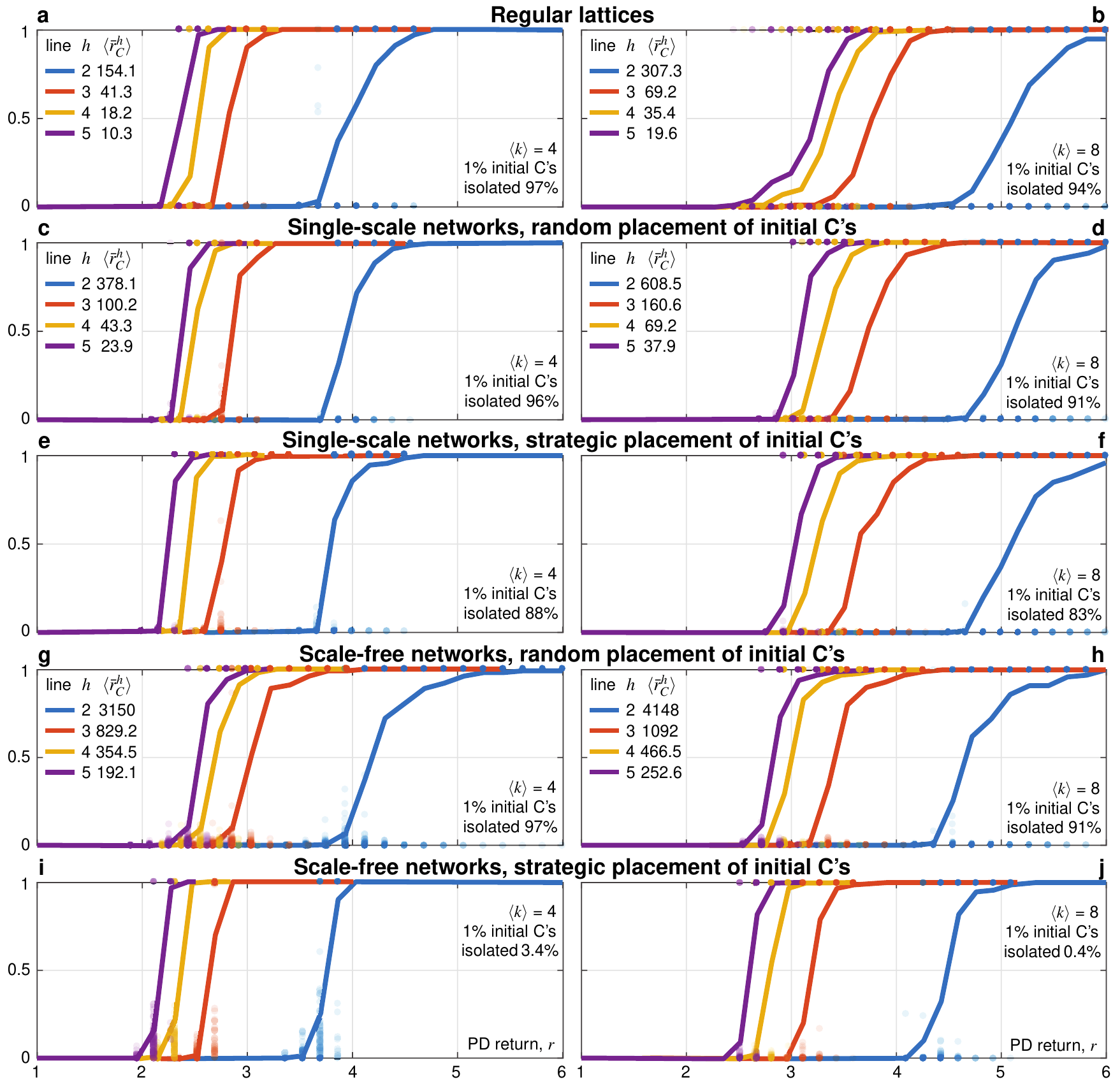}
\caption{Invasion, persistence, and fixation of cooperation under networked rational reciprocity.
Panels show the level of cooperation
reached in $10^4$ game rounds starting from $1\%$ inital C's on different network structures
(left: average degree $\langle k\rangle=4$; right: $\langle k\rangle=8$) as a function of the PD game return $r$.
Solid lines show the average fraction over $100$ random initializations
(random placement of the initial C's in planar lattices; network generation and random placement of the initial C's for random---single scale and scale free---networks; the average $\%$ of isolated initial C's is reported).
Dots show the outcome of single simulations
(only for $r_{\mathrm{C},\min}^h<r<r_{\mathrm{C},\max}^h$, i.e., only if some of the outcomes lie in the open interval $(0,1)$); 
transparency is used to show dots accumulation.
Colors code the predictive horizon $h$, from $2$ to $5$, and the corresponding upper bound $\langle\bar r_{\mathrm{C}}^h\rangle$ to the threshold $r_{\mathrm{C}}^h$ is reported.
See Sects.~\ref{sec:net} and~\ref{sec:num} for further details on networks and numerical simulations.
}
\label{fig:r01}
\end{figure}

To quantify the analytical results, we have run many numerical simulations on several networks of $N=1000$ nodes:
ring and planar lattices, single-scale (Watts-Strogatz model with full rewiring) and scale-free (Barabási-Albert model) random networks, and the complete network.
The results for 1\% initial fraction of (normally reciprocating, $\eps=0$) C's on planar $4$- and $8$-neighbor lattices and on random networks with average degree $\langle k\rangle=4$ and $8$ are reported in Fig.~\ref{fig:r01}.
See SI
Fig.~\ref{fig:r50} for 50\% initial C's and Fig.~\ref{fig:rcn} for a degree-$4$ ring lattice
(a ring of nodes each connected to the $4$ nearest nodes in the loop)
and the complete network;
see also Fig.~\ref{fig:red} for different values of the parameters $\eps$ and $\delta$.
For random networks, we have separately simulated the random placement of the initial C's and the placement according to degree-ranking.

The simulations confirm that networked rational reciprocity allows cooperation to invade, persist, and fixate in any network, provided the game return $r$ is large enough.
For any combination of network structure, initialization, and model parameters, the simulations starting from 1\% initial C's show two thresholds on $r$: an $r_{\mathrm{C},\min}^h$ above which, on average, cooperation invades and persists; and an $r_{\mathrm{C},\max}^h$ above which cooperation always invades and fixates.
For $r$ in between $r_{\mathrm{C},\min}^h$ and $r_{\mathrm{C},\max}^h$, the average asymptotic fraction of C's (solid lines) is not representative of the level of cooperation one should expect in a single simulation (dots), as cooperation disappears/fixates in most of the cases (see dots at fractions $0$ and $1$)
and typically settles at low C-levels (below $20\%$) in the rest of the cases.
A deeper analysis of the simulations of Fig.~\ref{fig:r01} not ended in all-C or all-D (the dots in the open interval $(0,1)$) indeed reveals that most of dots above $0.2$ denote simulations that converge to all-C on a longer timescale, whereas
dots below $0.2$ typically represents simulations ending in a nontrivial stalemate---different from all-C and all-D---or showing long-term fluctuations
(see SI
Sect.~\ref{sec:num} for further details and Sect.~\ref{sec:nf1} for examples of nontrivial stalemates and fluctuations in the simple network of Fig.~\ref{fig:net}{\sffamily\textbf b}).

Note that the threshold $r_{\mathrm{C},\max}^h$ is much smaller than the theoretical $r_{\mathrm{C}}^h$ discussed at point (viii) of the Analytical results
(see the average $\langle \bar r_{\mathrm{C}}^h\rangle$ over the simulations of a given type in Fig.~\ref{fig:r01}, where $\bar r_{\mathrm{C}}^h$ is the upper bound to $r_{\mathrm{C}}^h$ derived in
Sect.~\ref{sec:rch}).
This is due to the stochastic effect described at point (ix).
Essentially, even if some of the C's initially need a higher $r$ to remain C, by the time they revise their strategy the $r$-gap could have vanished because of D-to-C changes in the neighborhood.
This overcompensates the opposite effect due to the fact that $r_{\mathrm{C}}^h$ is computed starting from a cluster of two C's, whereas initial C's are most often isolated
(except for scale-free networks with degree-rank-C-placement, because hubs are likely connected among themselves;
see the average \% for each panel in Fig.~\ref{fig:r01}).
Starting with random pairs of connected initial C's indeed results in lower $r_{\mathrm{C},\min}^h$ and $r_{\mathrm{C},\max}^h$
(Fig.~\ref{fig:r2c}).

As expected from the arguments at points (x) and (xi), the thresholds $r_{\mathrm{C},\min}^h$ and $r_{\mathrm{C},\max}^h$ decrease with the horizon $h$ and increase with the network's average degree, given all other details
(Fig.~\ref{fig:r01}: compare the different colors within each panel and left vs right panels).
Note that the effects are weakened (as expected) starting from higher initial C-levels (see Fig.~\ref{fig:r50}).

Degree heterogeneity also works as theoretically predicted.
Placing the initial C's in the network's hubs does favor cooperation, both in terms of invasion and fixation
(Fig.~\ref{fig:r01}: compare single-scale and scale-free networks under degree-rank-C-placement and note the lower $r_{\mathrm{C},\min}^h$ and $r_{\mathrm{C},\max}^h$ in scale-free networks; also note that the type of placement is irrelevant for single-scale networks).
The effect is however moderate.
Degree heterogeneity works against the fixation of cooperation under random placement
(Fig.~\ref{fig:r01}: compare single-scale and scale-free networks and note the significantly higher $r_{\mathrm{C},\max}^h$ in the latters; $r_{\mathrm{C},\min}^h$ is slightly lower, because of the larger number of low-connected D's connected to the initial C's, who change to C under mild returns).
The effects are again weakened starting from higher initial C-levels (see Fig.~\ref{fig:r50}).


\section*{Discussion}
We have questioned network reciprocity as a mechanism supporting cooperation in socio-economic networks.
The imitation processes used for strategy update in all theoretical models lack sense when interaction patterns are local and/or heterogeneous, and the interest for unconditional cooperation disappears under any rational strategy update.
Imitation indeed did not emerge in social experiments
\citetex{Cassar_07_GEB,Kirchkamp_and_Nagel_07_GEB,Traulsen_et_al_10_PNAS,Grujic_et_al_10_PONE,gracia2012heterogeneous,Grujic_et_al_12_SR,Rand_et_al_14_PNAS,Li_et_al_18_PNAS}.
What did emerge \citetex{Grujic_et_al_10_PONE,gracia2012heterogeneous,gracia2012human,Grujic_et_al_14_SR} is a mood to be C or D and a form of direct reciprocity\citetex{Trivers_71_QRB,Axelrod81} in the C mood---the tendency to cooperate with subjects who showed cooperation in past interactions.
Nevertheless, the network plays a fundamental role in making cooperation possible. 
Static and sparse networks indeed make direct reciprocity feasible, granting repeated interactions within same groups of individuals and limiting the required cognitive abilities.
We have theoretically shown that these are the networks in which a simple form of direct reciprocity, driven by a rational strategy update, more easily allows cooperation to invade and reach fixation.

\subsection*{Aims and novelty of the model}
\label{sec:nov}
Our primary motivation was not to model the human behavior observed in the available experiments, but rather to test a specific hypothesis:
Whether a form of direct reciprocity could evolve on networks (invasion and fixation) under a process of strategy update that rationally pursues the individual's interest.
We named this EGT scenario {\it networked rational reciprocity}, of which our model is a minimal benchmark version.
Minimal in two respects: the number of confronted strategies---two, strategies C and D---and, especially, the amount of information available to the players---nothing more than what they know from direct interaction.
Benchmark, because of the choices of the PD interaction and the D strategy---unconditional D---both setting the worst-case scenario for the evolution of cooperation.

The main novelty of our model is the model-predictive rule for strategy update.
Except for best-response update\citetex{Roca2009}---corresponding to our 1-step prediction---the evolution of unconditional C on networks has been always studied with update rules that implement imitation in socio-economic contexts.
Also direct reciprocity has been similarly investigated. \citetex{Ohtsuki_and_Nowak_07_JTB}

Other elements are rather standard. \citetex{Szabo_and_Fath_07_PR}
Individuals repeatedly play in a static network.
At each round, each individual is in the C or D mood\citetex{gracia2012human,Grujic_et_al_14_SR} and accordingly behaves using the C or D strategy.
A round consists of a PD interaction among all pairs of neighbors and payoffs are collected.
Strategy update is slow---compared to the frequency of game rounds---and asynchronous: \citetex{Grilo_and_Correia_12_JTB}
each individual revises what is the best strategy to follow at a rate that is assumed small and uniform across the population.
This is the assumption that makes predictions of short-future incomes possible, by disregarding the neighbors' updates in the predictive horizon.

The way in which we implement direct reciprocity---allowing C's to abstain from playing---is also not new.
Optional participation is known to relax social dilemmas when a baseline payoff is granted to loners,
\citetex{Hauert_et_al_02_SCI,Szabo_and_Hauert_02_PRE}
whereas link disconnection has been recently considered. \citetex{Li_et_al_17_SR}
Our link inhibition is temporary and grants no profit to C's.
We prefer abstention rather than forcing retaliation---cooperators defecting neighboring exploiters---because this is more connatural to the C mood.
Although there is no difference in a single round (because we assume no payoff for mutual defection), abstaining C's communicate their mood to exploiters.
This choice is however not crucial for our findings. 
We expect similar results by confronting unconditional D with any reciprocating form of conditional cooperation under a rational process of strategy update
(e.g., we preliminary tested the well-known tit-for-tat and forgiving tit-for-tat\citetex{Axelrod81}).

Direct reciprocity deserves another comment.
Originally,\citetex{Trivers_71_QRB,Axelrod81} it has been studied in iterated games, i.e., (non-evolutionary) games involving only two players (rather than a population of two types of players) who know the probability $w>0$ of a next interaction.
In the evolutionary context, there are two ways in which one can study repeated interactions among two given players in the population.
Either the single game round consists of an iterated game between each pair of neighbors---an option allowing direct reciprocity even in large, dense or highly-dynamic networks, so far investigated in the socio-economic context under imitation-like update rules\citetex{Ohtsuki_and_Nowak_07_JTB}---or one relies on a static, sparse network, as we do.
Our game round involves a one-shot, optional PD interaction with neighbors.
The game is however repeated over an indefinite number of rounds in a static network, so that a next round is essentially always granted.
The sparsity of interactions makes direct reciprocity possible, by limiting the cognitive abilities required to remember neighbors and past interactions.
Of course hubs need more resources than leaves, but this is typically built-in in the socio-economic structure.

\subsection*{From network reciprocity to networked rational reciprocity}
\label{sec:nrr}
Network reciprocity and networked rational reciprocity are substantially different mechanisms for the evolution of cooperation.
Limiting the discussion to socio-economic systems, the former models the competition between unconditional cooperation and defection under an imitation-like process of strategy update;
the latter studies the competition between a reciprocating form of conditional cooperation and unconditional defection under a rational strategy update.
Their common bond is the need of a population structure, steady and local, to support cooperation.
To allow cluster of C's protected from D's and to make direct reciprocity feasible and effective, respectively.

We confirm the fundamental role played by static and sparse networks of contacts in the evolution of cooperation.
We rethink, however, the underlying evolutionary mechanism.
Direct reciprocity combined with a farsighted rule of strategy update---our multi-step predictive horizon---are the keys to explain the success of cooperation.
With no reciprocity (or other mechanisms) supporting cooperation, players should rationally opt for defection;
this is well-known in unstructured population, but it works as well on any structure.
Similarly, the myopic optimization of the next game round suggests to defect.

Interestingly, our results are in line with those theoretically obtained for network reciprocity:
\citetex{Ohtsuki_et_al_06_NAT,Konno_11_JTB,Allen_et_al_17_NAT}
there are good chances that cooperation invades and fixates if the game return $r$ sufficiently exceeds a measure of the network connectivity (see condition~\eqref{eq:condrinf} for the case of an infinite predictive horizon;
the threshold $r_{\mathrm{C}}^h$ in Sect.~\ref{sec:rch} for a finite horizon;
also see all our numerical simulations).
Two aspects on which the evolution of cooperation under network reciprocity and networked rational reciprocity differ are discussed in the following two sections.

\subsection*{The invasion of cooperation}
\label{sec:inv}
We have shown that if the game return $r$ is large enough, networked rational reciprocity grants high chances of (invasion and) fixation starting from a single C, i.e., chances of the order $1-1/(k+1) + O(\delta)$, where $k$ is the degree of the initial C and $\delta$ is the rate of strategy update (see analytical results (viii) and (ix)).
This is different from what is granted by network reciprocity under the rule $r>k$, i.e., fixation probability larger than $1/N$
(in large regular networks in the limit of weak selection;
$1/N$ is the fixation probability under totally random strategy update).
When the rule of network reciprocity is weakly satisfied in a large network, cooperation almost surely disappears starting from a single C (probability $1-1/N$).
To have higher chances of fixation, a significantly larger $r$ is typically required and, especially when selection is strong (SI note 3), cooperation cannot invade anyhow.
Consider, e.g., a single C in the lattice of Fig.~\ref{fig:net}{\sffamily\textbf a}.
If selection is strong, the C most likely imitates a D-neighbor as soon as she revises her strategy (probability $\delta$), whereas D-neighbors do not imitate the C.
The probability of invasion---to go from one to two C's---is negligible after each game round, whereas the C sooner or later switches to D.

We hence conclude that network reciprocity does not support the invasion of cooperation.
Not surprisingly, all theoretical studies showing significant fixation probabilities for cooperation under intermediate or strong selection considered random initial conditions with 50\% C's
\citetex{Santos_and_Pacheco_05_PRL, Santos_and_Pacheco_06_JEB, Santos_et_al_06_PNAS, Santos_et_al_06_PRSB, Gomez-Gardenes_et_al_07_PRL, Poncela_at_al_07_NJP, Assenza_et_al_08_PRE, Gomez-Gardenes_et_al_08_JTB, Pusch_and_Weber_08_PRE, Santos_et_al_08_NAT, Devlin_and_Treloar_09_PRE_a, Devlin_and_Treloar_09_PRE_b, Floria_et_al_09_PRE, Perc_09_NJP, Cardillo_et_al_10_NJP, Debarre_et_al_14_NCOM}
($33\%$ has been considered in Ref.~\onlinecite{Roca2009}).
Starting from isolated C's or small clusters, cooperation most likely disappear.
We have, e.g., tested network reciprocity starting from 1\% initial C's on the same network structures of Fig.~\ref{fig:r01}, using the pairwise comparison imitation rule under strong selection (the one used in the majority of the above mentioned works; see SI notes 2 and 3).
Cooperation systematically disappeared up to $r=5000$, except for scale-free networks with degree-rank-C-placement
(where C-hubs are known to form clusters)
in which we found invasion only for $r$ larger than $20$.

\subsection*{The effect of the network structure}
\label{sec:str}
A considerable effort has been devoted to identify the network structures that best favor the evolution of cooperation under network reciprocity.
\citetex{Santos_and_Pacheco_05_PRL, Santos_and_Pacheco_06_JEB, Santos_et_al_06_PNAS, Gomez-Gardenes_et_al_07_PRL}
The general answer is that, for a given (sufficiently small) average degree $\langle k\rangle$ and starting from a significant C-level, heterogeneous networks---e.g., scale-free networks---do better than homogeneous networks---lattices or single-scale networks.
Indeed, a C-hub (individual $i$ with degree $k_i\gg\langle k\rangle$) with a significant fraction (say 50\%) of C-neighbors is imitated by a low-connected D-neighbor $j$ under a mild requirement on the game return $r$
($\pi_i = (r-1)k_i/2-k_i$, $\pi_j=r$ for a leaf $j$).
C-hubs can then build C-clusters, whereas this requires higher returns in homogeneous networks ($k_i,k_j\approx\langle k\rangle$).

On the contrary, heterogeneity works against cooperation under networked rational reciprocity.
From condition~\eqref{eq:condrinf} 
(for the case of an infinite horizon; similarly see the threshold $r_{\mathrm{C}}^h$ in Sect.~\ref{sec:rch} for a finite horizon)
we see that the threshold on $r$ above which strategy revising C's (resp.~D's) remain (resp.~change to) C increases with the node degree $k$.
Especially at low levels of cooperation (low $k_{\mathrm{C}}$), C- or D-hubs (connected to one or a few C's) require a larger $r$ to opt for C than nodes with degree closer to average.
The simulations in Fig.~\ref{fig:r01} indeed show that the $r$-thresholds required for invasion and fixation of cooperation are higher for scale-free w.r.t. single-scale networks, if the initial C's are placed at random.

However, C-hubs encourage low-connected D-neighbors in changing to C under mild returns.
Provided the rate of strategy update is small enough, the number of C-neighbors of a C-hub will raise, while the hub pays the cost of building the cluster.
In other words, initially isolated C-hubs invest in the future establishment of cooperation.
Moreover, hubs are likely connected among themselves, so that placing the initial C's in the network's hubs forms clusters of C's that mitigate the investment.
Network heterogeneity therefore turns beneficial to cooperation under this strategic placement of the initial C's.

\subsection*{Links with social experiments}
\label{sec:exp}
Four experiments on relatively large ($N\ge 100$), static and sparse human networks playing a PD have been performed to date.
\citetex{Grujic_et_al_10_PONE,gracia2012heterogeneous,Rand_et_al_14_PNAS,Li_et_al_18_PNAS}
The two most recent
(on ring lattices: $N=100$, degree $k=2,4,6$\citetex{Rand_et_al_14_PNAS}; $N=225$, $k=2$\citetex{Li_et_al_18_PNAS})
have shown significant levels of stable cooperation under the rule $r>k$ of network reciprocity
($r=2,4,6$\citetex{Rand_et_al_14_PNAS} and $r=2$\citetex{Li_et_al_18_PNAS}; significant cooperation was observed also for $r=k$),
though the consistency of human behavior with the unconditional C and D strategies under imitation update was not documented.
In the first two experiments
(on planar lattices: $N=13\times 13=169$, $k=8$\citetex{Grujic_et_al_10_PONE}; $N=25\times 25=625$, $k=4$\citetex{gracia2012heterogeneous};
on a heterogeneous network: $N=604$, degree from $2$ to $16$, $\langle k\rangle=3.4$\citetex{gracia2012heterogeneous}),
the game return $r$ was set below ($r=10/3$) the average connectivity and cooperation dropped at levels comparable to those expected in the complete network (the all-to-all interaction was mimicked in a control treatment by reshuffling neighbors at each round).
Subjects were however documented not to imitate better performing neighbors.
They cooperated by essentially reciprocating the benefit obtained in the previous round.
Moreover, the positive effect predicted by network reciprocity in heterogeneous networks
\citetex{Santos_and_Pacheco_05_PRL, Santos_and_Pacheco_06_JEB, Santos_et_al_06_PNAS, Gomez-Gardenes_et_al_07_PRL}
was missed.\citetex{gracia2012heterogeneous,gracia2012human}
Because the experimental setting was very similar in the four experiments---in particular, the same game option, C or D, is taken for all neighbors at each round---we might expect a similar behavior.

Other experiments have been performed on smaller networks, with results consistent with the four larger experiments.
On small ring lattices ($N=18$, $k=4$), low/medium levels of cooperation was observed for $r=4,5$,
\citetex{Cassar_07_GEB,Kirchkamp_and_Nagel_07_GEB}
with subjects reacting to cooperation in previous rounds rather than to neighbors' previous payoffs
(the payoff per round was however normalized by the node degree).
Small-words and random networks ($N=18$, $k=4$)\citetex{Cassar_07_GEB} as well as complete networks ($N$ from $2$ to $5$)
\citetex{Kirchkamp_and_Nagel_07_GEB,Grujic_et_al_12_SR}
showed lower cooperation w.r.t. homogeneous sparse networks.
On small square lattices ($N=16$, $k=4$), low cooperation was observed for $r=3$, \citetex{Traulsen_et_al_10_PNAS}
with an apparently unconditional behavior driven by imitation update, though the cooperative strategy was later shown more robustly described as conditioned by direct reciprocity under a random strategy update \citetex{Grujic_et_al_14_SR} (see below).

The identification of the strategy (one or many) adopted (or learned) by humans when playing a PD experiment is a difficult task and the result is likely to depend on the experimental setting.
A few general traits are however apparent from the analysis of the experiments of Refs.~\onlinecite{Grujic_et_al_10_PONE,Traulsen_et_al_10_PNAS,gracia2012heterogeneous} (with similar settings).
Apart from subjects who mostly cooperate or defect, that are always minorities, the analysis revealed that humans behave according to a cooperative or defective mood, \citetex{gracia2012human,Grujic_et_al_14_SR}
thus justifying the modeling assumption of two strategies.
In the C mood, subjects more likely cooperate (with all neighbors, as a rule of the experiment) the more cooperation was observed in the neighborhood in the previous round, i.e., they reciprocate cooperation.
Subjects in the D mood most likely defect irrespectively of the previous round, i.e. defection is largely unconditional.
Strategy change is identified as a random process biased by the subject's mood.
Precisely, a C/D-player changes strategy after opting for defection/cooperation at a given round.
This shows that subjects did not copied better performing neighbors (though payoff were made available), but does not unveil what pushes subjects to change strategy.
Our basic rational hypothesis is that subjects try to maximize, at the best of their knowledge, the outcome of one or a few future rounds.

Summarizing, the social experiments show that the C strategy is conditioned by direct reciprocity,
that defection is rather unconditional, and that
a rule of the kind $r>\langle k\rangle$ essentially works---in the sense of a threshold on $r$ over which cooperation stabilizes at levels that increase with the game return $r$.
They do not suggest the underlying evolutionary mechanism, but show it is not network reciprocity.
These results provide the motivation and the basis for our model.
However, we imagined a different model society with respect to the one imposed by the experimental setting in
Refs.~\onlinecite{Grujic_et_al_10_PONE,Traulsen_et_al_10_PNAS,gracia2012heterogeneous}.
The two major differences concern the C strategy.
We assume that C-players do not defect,
rather they can abstain from playing with defecting neighbors;
and the choice is taken independently for each neighbors, instead of forcing a common decision.
Assuming a rational process of strategy update, based on a model prediction of future payoffs, we show that cooperation fixates if the game return $r$ is larger than a threshold that scales with the average connectivity of the network.

Although our simulations cannot be directly compared with the available experiments,
we provide an evolutionary mechanism that has the potential to explain human cooperation.
Our model-predictive rule for strategy update is not an easy one to apply in real networks, in which the rate of strategy update might be far from uniform across the population.
Moreover, it requires nontrivial cognitive abilities and humans might not be as rational as we assume.
However, it could emerge, approximatively, as the result of an intuitive, rather than computational, human behavior.
To test this claim new experiments must be designed, and our model suggests how.
For example, to confirm that humans do have a C or D mood, it is important to allow them to temporarily abstain from playing with specific neighbors, to avoid confounding risk-avoiding defections with a D mood.
Allowing independent decisions with different neighbors is also important, e.g., to identify a mix of C and D as the absence of a mood.
Allowing independent decisions definitely poses experimental challenges, and it has been recently shown to enhance cooperation in static networks.
\citetex{Melamed_et_al_18_PNAS}

\section*{Methods}
When the C-individual $i$ (normally reciprocating, $\eps=0$) gets exploited by neighbor $j$, she draws the number $a$ of game rounds to skip according to the distribution $\text{Prob}(a)=(1-\delta)^{a}\delta$, $a\ge 0$ (with mean $1/\delta-1$), i.e., $i$ goes back playing with $j$ at the $(a+1)$-th round following the exploitation with the probability that $j$ first revises
her strategy after the $a$-th round
(e.g., $i$ does not stop playing with $j$, $a=0$, with the probability $\delta$ that $j$ revises strategy just after exploiting $i$).
Without knowing the number drawn by $i$, the probability $p_{ij}$ that $i$ will agree to play
with $j$ at the $t$-th round following exploitation is the cumulative distribution of $\text{Prob}(a)$ from $a=0$ to $a=t-1$, i.e., the probability\\[-2mm]
\begin{equation}
\label{eq:pt}
p_t=1-(1-\delta)^t
\end{equation}
that the drawn number is
smaller than $t$.

Instead of managing abstention periods, we implement our model by endowing each individual $i$ with the set of probabilities $p_{ij}$ that $i$ will agree to play
with $j$ at the next round, $i,j=1,\ldots,N$.
Initially, $p_{ij}=p_{ji}=1$ if $i$ and $j$ are neighbors; $p_{ij}=p_{ji}=0$ otherwise.
The $N\times N$ matrix $P=[p_{ij}]$ defines the (static) network topology.
At each game round, each PD interaction takes place with probability $p_{ij}p_{ji}$, i.e., only if both players agree to play.
When the C-individual $i$ gets exploited by neighbor $j$, she sets $p_{ij}=p_1=\delta$, i.e., to the probability that $j$ revises
her strategy just after.
If $i$ decides not to play with $j$ at the next round, $p_{ij}$ is updated to $p_2=1-(1-\delta)^2$, the probability that $j$ had revised strategy by the second round following exploitation.
After $t-1$ consecutive abstentions, the probability to play at the $t$-th round is $p_t$ in \eqref{eq:pt}, that increases to one with $t$.
When the C-individual $i$ gets reciprocated by the C-neighbor $j$, she resets $p_{ij}=1$.
D's always have $p_{ij}=1$ toward all neighbors.
(Note that either $p_{ij}=1$ or $p_{ji}=1$ by construction.)
We therefore implement a different model, w.r.t. the one so far described, by shifting the network's dynamic from the links (on-off) to the links' weights (the probabilities $p_{ij}$).
The two models are statistically equivalent and the latter is simpler to analyze
(see Supplementary Information).

For super/sub reciprocating C's ($\eps\gtrless 0$), the rate of strategy update $\delta$ must be replaced with the biased rate $\delta_{\eps}$ in the above formulas.
Recall (from~\eqref{eq:deps}) that $\eps=-(\delta_{\eps}-\delta)/\delta$ is the relative mismatch between $\delta$ and $\delta_{\eps}$, i.e., the under/over-estimation of $\delta$ adopted by super/sub reciprocating C's in deciding how long to abstain.
For super/sub reciprocating C's, $1/\delta_{\eps}$ is larger/smaller than the average number of rounds $1/\delta$ within which the exploiter once revises her strategy, so that, on average, C's stop playing with exploiters for longer/shorter than the time taken by the latter to possibly change to C.
(C's stop playing forever if $\eps\to\eps_{\max}=1$; they never stop---they play C unconditionally---if $\eps\to\eps_{\min}=-(1-\delta)/\delta$.)

After each game round, each individual independently decides whether to revise strategy with probability $\delta$.
C's who revise compute their expected accumulated payoff assuming to remain C, $\pi_{\mathrm{CC}}^h$, or to behave as D, $\pi_{\mathrm{CD}}^h$, in the next $h$ rounds
(see
Sect.~\ref{sec:DpiC})
and change to D if the expected gain $\Delta\pi_{\mathrm{C}}^h=\pi_{\mathrm{CD}}^h-\pi_{\mathrm{CC}}^h$ is positive.
So doing, they assume the neighbors' strategy unchanged since the last PD interaction that took place, so that D's who changed to C in the meantime are considered as D's.
D's who revise compute their expected accumulated payoff gain $\Delta\pi_\mathrm{D}^h=\pi_{\mathrm{DC}}^h-\pi_{\mathrm{DD}}^h$ under full information
(see
Sect.~\ref{sec:DpiD})
and change to C under a positive gain.
When changing to D, C's set $p_{ij}=1$ toward all neighbors.
When changing to C, D's set $p_{ij}=\delta_{\eps}$ toward D-neighbors, as if drawing an abstention period.

To compute their expected payoff gains $\Delta\pi_{\mathrm{C}}^h$ and $\Delta\pi_\mathrm{D}^h$, C's and D's evaluate the probabilities to interact with their neighbors during the predictive horizon.
To this end, two probabilities are defined in
Sects.~\ref{sec:pCDh} and~\ref{sec:pCCh} for the C-individual $i$ with $p_{ij}=p_{t_{ij}}$ for some $t_{ij}\ge 0$
($p_0=1$ by definition):
the probability $P_{\mathrm{CD}}^{t}(t_{ij})$ to play with the D-neighbor $j$ at round $t\ge 1$ of the horizon;
and, similarly, the probability $P_{\mathrm{CC}}^{t}(t_{ij})$ to play with the C-neighbor $j$ (with $p_{ji}=1$).
As $t\to\infty$, $P_{\mathrm{CD}}^{t}$ converges (independently of the initialization) to the infinite-horizon limit $P_{\mathrm{CD}}^{\infty}$ of condition~\eqref{eq:condrinf}.
Although making prediction assuming no strategy change in the neighborhood makes sense only for short horizons, relatively to the revision rate $\delta$, players can hardly do better predictions anyway.
In our simulations we have limited the product $\delta h$---upper bounding the neighborhood fraction possibly subject to change within the horizon---to $0.3$ (e.g., $\delta=0.05$ and $h\le 5$ in Fig.~\ref{fig:r01}).

For details on networks' structure and generation and on numerical simulations, see
Sects.~\ref{sec:net} and~\ref{sec:num}, respectively.

%
\vspace{4mm}\noindent{\sffamily\textbf{Author contributions statement:}}
All authors were involved in the design of the research and in the analysis. 
F.D.R. performed the numerical analysis and produced the graphics;
C.P. performed the network analysis;
F.D. wrote the paper.

%

\bibliographystyletex{naturemag}
\bibliographytex{egtnetbib}
\label{finalpagearticle}

\clearpage

\renewcommand{\theequation}{S\arabic{equation}}
\setcounter{equation}{0}
\renewcommand{\thefigure}{S\arabic{figure}}
\setcounter{figure}{0}
\renewcommand{\thetable}{S\arabic{table}}
\setcounter{table}{0}
\renewcommand{\thesection}{S\arabic{section}}
\setcounter{section}{0}
\renewcommand{\thesubsection}{S\arabic{subsection}}
\setcounter{section}{0}

\setcounter{page}{1}
\rfoot{\small\sffamily\bfseries\thepage/\pageref{LastPage}}%

\makeSItitle

\addtolength\textheight{12pt}

\section*{Supplementary Notes}
\vspace*{5mm}
\begin{enumerate}
\item
We focus on static networks.
Dynamical networks, where old contacts can be broken and new ones established---exogenously or in feedback from the game interaction---and whether this favors cooperation is not discussed in this work.
\citeSI{Pacheco_et_al_06_PRL,Pacheco_et_al_06_JTB,Rand_et_al_11_PNAS,cuest2015reputation}
\item
Network reciprocity has been shown to work under several evolutionary processes.
These include \citetex{Ohtsuki_et_al_06_NAT,Santos_et_al_06_PNAS} the two imitation processes traditionally used to describe strategy update in socio-economic networks:
{\it Imitation} (IM)---an individual is selected uniformly at random to revise her strategy and stays or copies one of the neighbors' strategies with probabilities proportional to her own and the neighbors' (normalized) fitnesses---and {\it Pairwise Comparison} (PC)---a randomly chosen individual compares payoffs with a random neighbor and stays or copies the neighbor's strategy proportionally to fitness difference.
Fitness is a measure of performance in the underlying game (in terms of reproduction in biology and status in socio-economic systems);
it can simply be the game payoff in the last round (see note~3).
Network reciprocity best works for the biological process known as {\it Death-Birth} (DB)\citetex{Ohtsuki_et_al_06_NAT}---a random individual is selected to die and the neighbors compete for the empty site proportionally to fitness---that is equivalent to a modified IM process where the selected individual is forced to imitate a neighbor.
Interestingly, network reciprocity does not work
(the condition on the game return $r$ is highly demanding \citetex{Allen_et_al_17_NAT})
for the dual biological process of {\it Birth-Death} (BD)---an individual is selected to reproduce proportionally to fitness over the whole population and the offspring replaces a randomly selected neighbor.
Essentially, under BD, D's at the border of C-clusters reproduce more than bordering C's.
(See Ref.~\onlinecite{Debarre_et_al_14_NCOM} for other biologically-inspired evolutionary processes, in which selection acts globally or locally on both birth and death with possibly independent dispersal and interaction graphs.)
\item
Weak selection means that the game payoffs contribute to the individual's fitness only to a small extent.
It is an interesting limit because it simplifies analytic computations.
It represents situations in which the individual's performance is largely determined by factors that are independent of the game interaction and typically assumed time-invariant.
They give a baseline fitness to all players, to which the game output marginally adds.
E.g., in the PC process, the probability that an individual with low payoff imitates a selected one with high payoff is slightly above 50\%;
as well, there is a significant probability, slightly below 50\%, to copy individuals with lower payoff.
That is, the performance in the game is weakly selected.
Selection is strong when the fitness is totally determined by the game. It can simply be the payoff, or even a function of the payoff that increases more than linearly, so to increase the probability that a given payoff difference will result in the selection of the best performance.
The strongest selection (extreme selection \citeSI{Della_Rossa_17_BMB}) is the case in which the player with larger payoff is always selected to reproduce or be copied.
In our model selection is extreme, in the sense that when revising strategy individuals always opt for the strategy giving the largest payoff prediction.
\item
The {\it transitivity} of a network---the average probability that the neighbors of a node are neighbors themselves---is best known as the network's {\it clustering}. \citeSI{BoLa:06} We use transitivity to avoid confusion with a C-cluster---a maximal group of connected C's.
\end{enumerate}

\newpage
\section*{Supplementary Figures}
\vspace*{5mm}

\begin{figure}[h]
\centering
\includegraphics{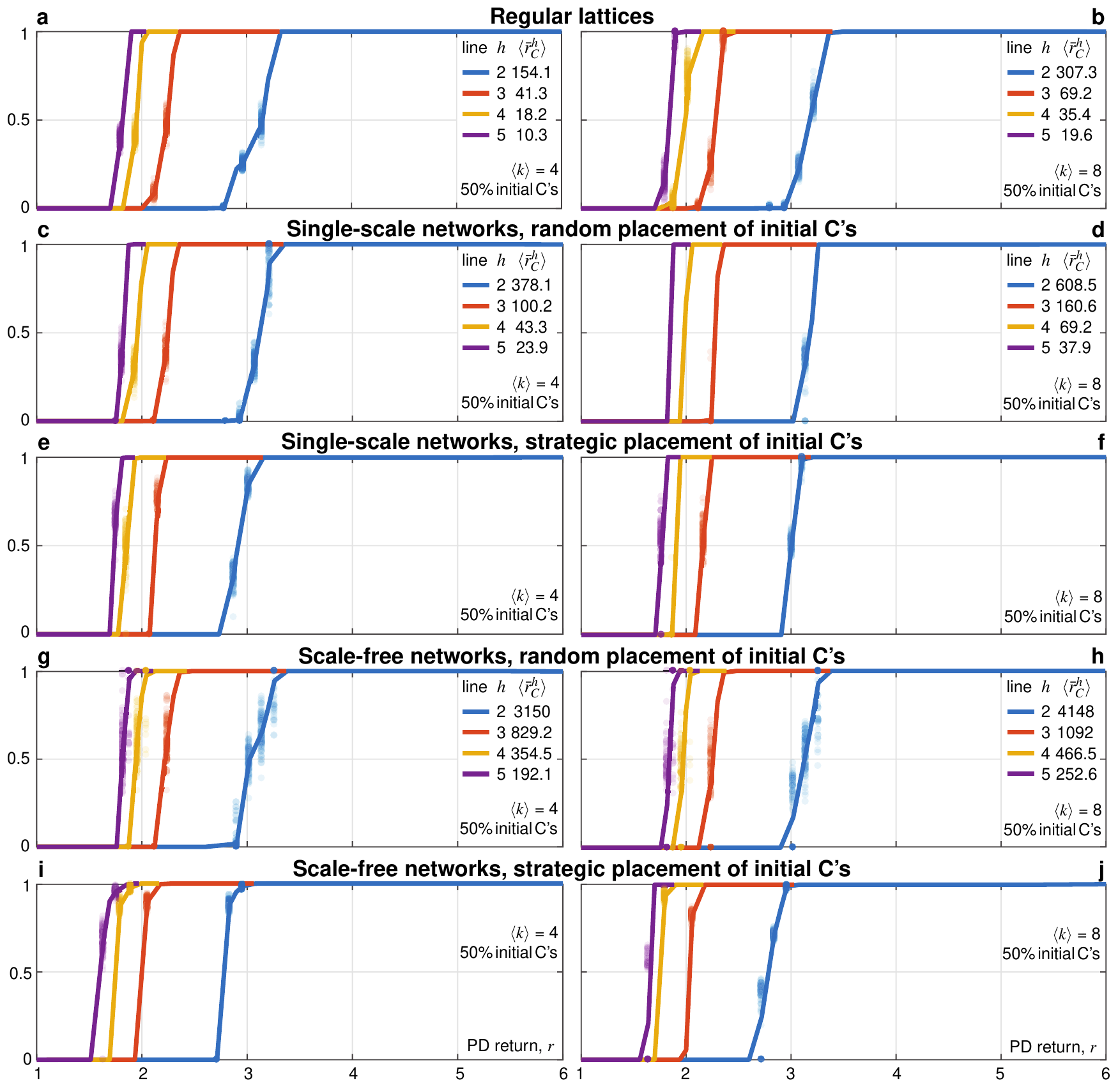}
\caption{Persistence and fixation of cooperation under networked rational reciprocity.
The figure complements Fig.~\ref{fig:r01} by showing the results obtained starting from $50\%$ initial C's, all other details unchanged (same model parameters, networks, and initializations).
Note that the thresholds on $r$ identified with $50\%$ initial C's have a different meaning w.r.t. the significantly higher $r_{\mathrm{C},\min}^h$ and $r_{\mathrm{C},\max}^h$ of Fig.~\ref{fig:r01}.
Here cooperation persists (resp. fixates), on average, for $r$ above the lower (resp. higher) threshold, being however unable to invade (resp. fixate after invasion) if $r$ is below the $r_{\mathrm{C},\min}^h$ (resp. $r_{\mathrm{C},\max}^h$) of Fig.~\ref{fig:r01}.
The effects of the different network structures and initializations observed in Fig.~\ref{fig:r01} are still present but weakened.}
\label{fig:r50}
\vspace*{-25mm}
\end{figure}

\begin{figure}[h]
\centering
\includegraphics{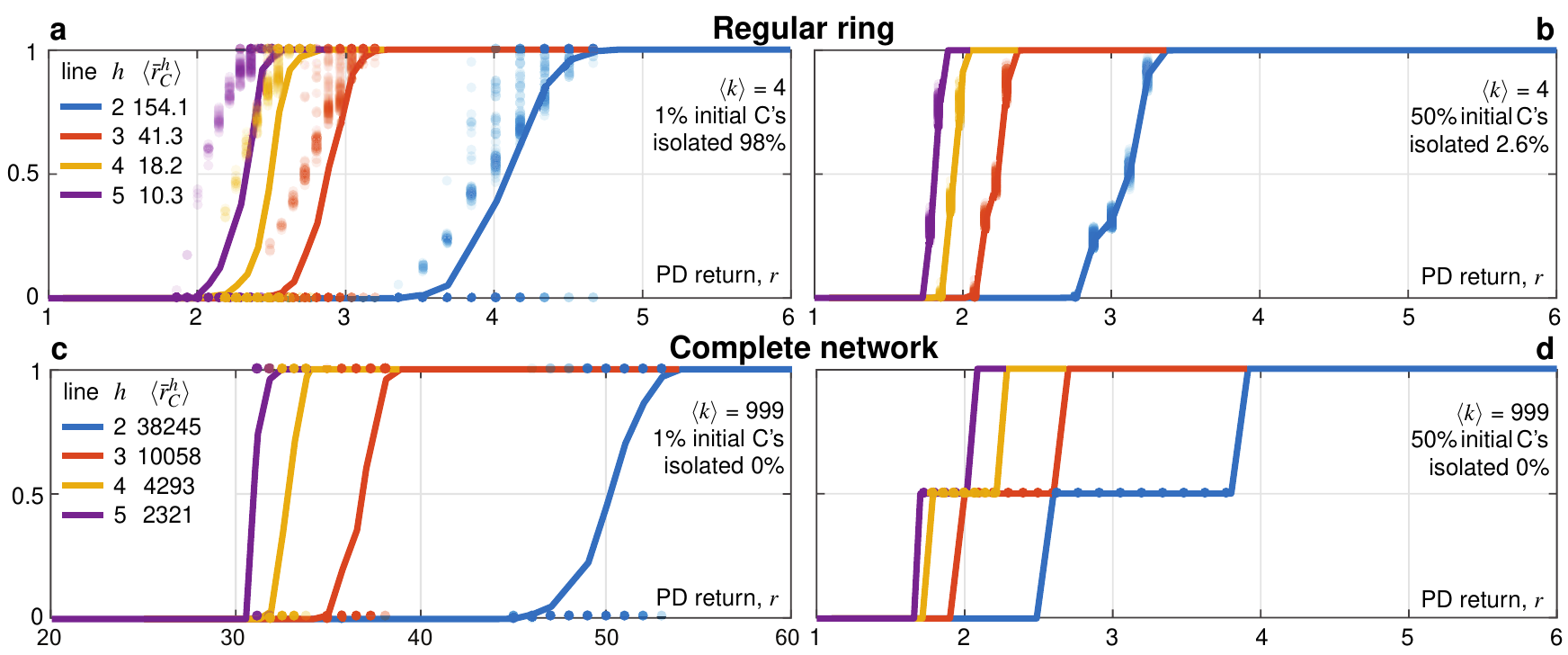}
\caption{Networked rational reciprocity on other regular network structures.
The figure complements Figs.~\ref{fig:r01} and~\ref{fig:r50} by adding a degree-$4$ ring lattice and the complete network,
starting from $1\%$ (left panels) and $50\%$ (right panels) initial C's.
The ring behaves similarly to the square lattice, better supporting the invasion of C's (slightly lower $r_{\mathrm{C},\min}^h$ for $1\%$ initial C's).
In the ring lattice, half of the D's at the boundary of a small C-cluster (those next to a C in the main loop) have $2$ D-neighbors, whereas they typically have $3$ D-neighbors in the square lattice, so the that D-to-C strategy changes requires a lower return at low levels of C.
The increased number of simulations not ended in all-C or all-D w.r.t. the square lattice
(dots in the open interval $(0,1)$, especially for $1\%$ initial C's) is due to the slower convergence, because of the essentially one-dimensional ring structure.
The all-to-all connection is the most demanding structure for the invasion of C's
(note the different scale of the $r$-axis in panel {\sffamily\textbf c}),
and even for its fixation starting from $50\%$ initial C's (compare with Fig.~\ref{fig:r50}).
The complete network is in a stalemate starting from $50\%$ initial C's with intermediate returns.
Stalemate at the initial state is possible on any network structure
(see, e.g., the simple network of Fig.~\ref{fig:net}{\sffamily\textbf b} analyzed in Sect.~\ref{sec:nf1}), though we observed it in our numerical simulations only for the complete network.
It typically requires significant C-levels, since isolated C's are willing to change, though we observed it also at $1\%$ initial C's for $r$ slightly larger than $r_{\mathrm{C},\min}^h$.
(See Sect.~\ref{sec:num} for more details on the numerical results.)
}
\label{fig:rcn}
\end{figure}

\begin{figure}[h]
\centering
\includegraphics{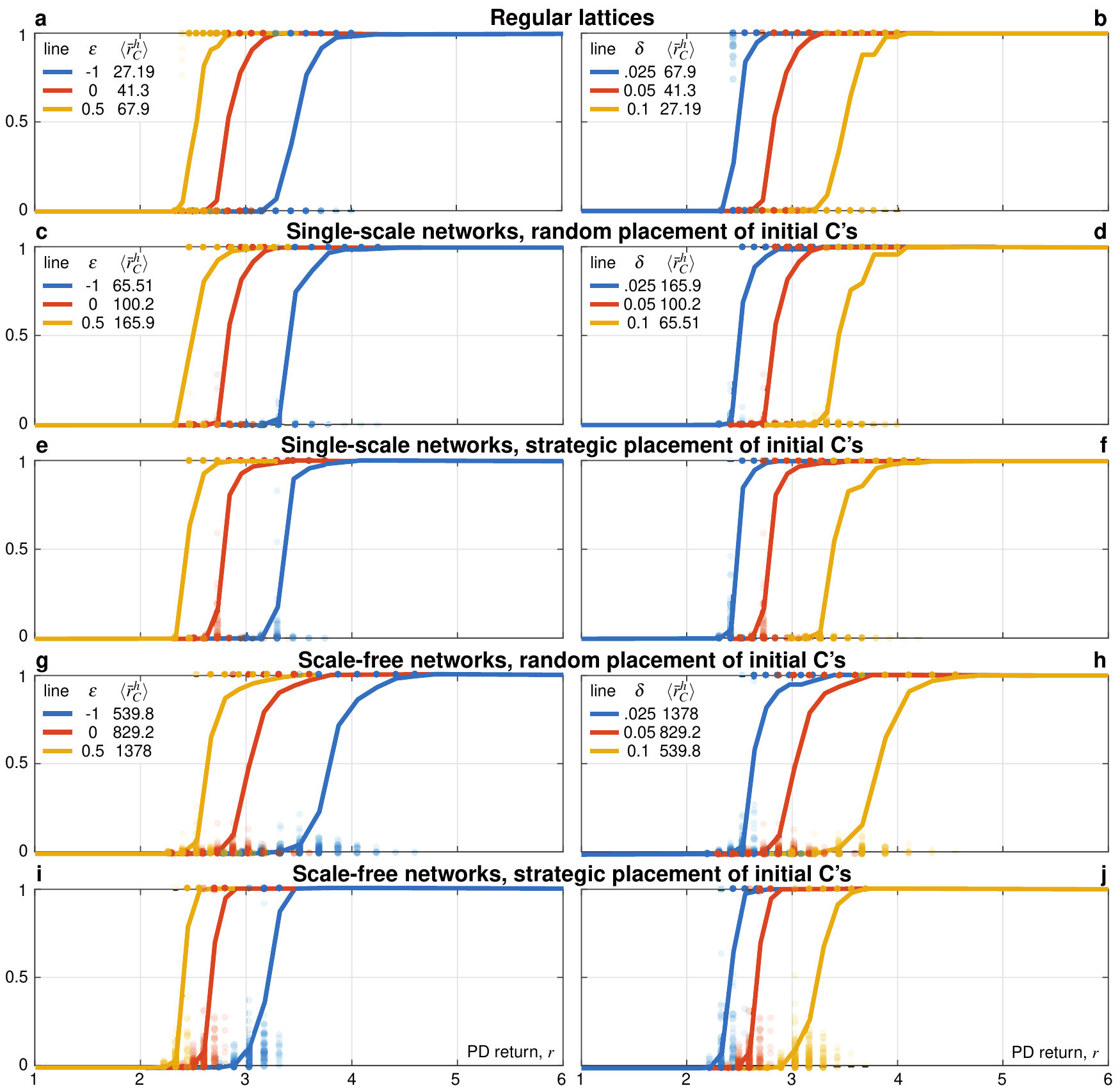}
\caption{Sensitivity analysis.
The figure complements Fig.~\ref{fig:r01} by reproducing the simulations with average degree $\langle k\rangle=4$ and predictive horizon $h=3$
(red in the left panels of Fig.~\ref{fig:r01}) for two different (larger/smaller) values of the reciprocity parameter $\eps$
($0.5$ and $-1$, halving/doubling the reciprocity-biased rate of strategy update $\delta_{\eps}$ (see~\eqref{eq:deps}) w.r.t. the baseline value in Table~\ref{tab:par}; left column)
and of the rate of strategy update $\delta$
($0.1$ and $0.025$, double/half of the baseline value; right column).
As expected at points (v) and (vii) of the Analytical results, super/sub reciprocation ($\eps\gtrless 0$) as well as higher/lower inertia to change (lower/higher $\delta$) result in lower/higher thresholds $r_{\mathrm{C},\min}^h$ and $r_{\mathrm{C},\max}^h$.}
\label{fig:red}
\end{figure}

\begin{figure}[h]
\centering
\includegraphics{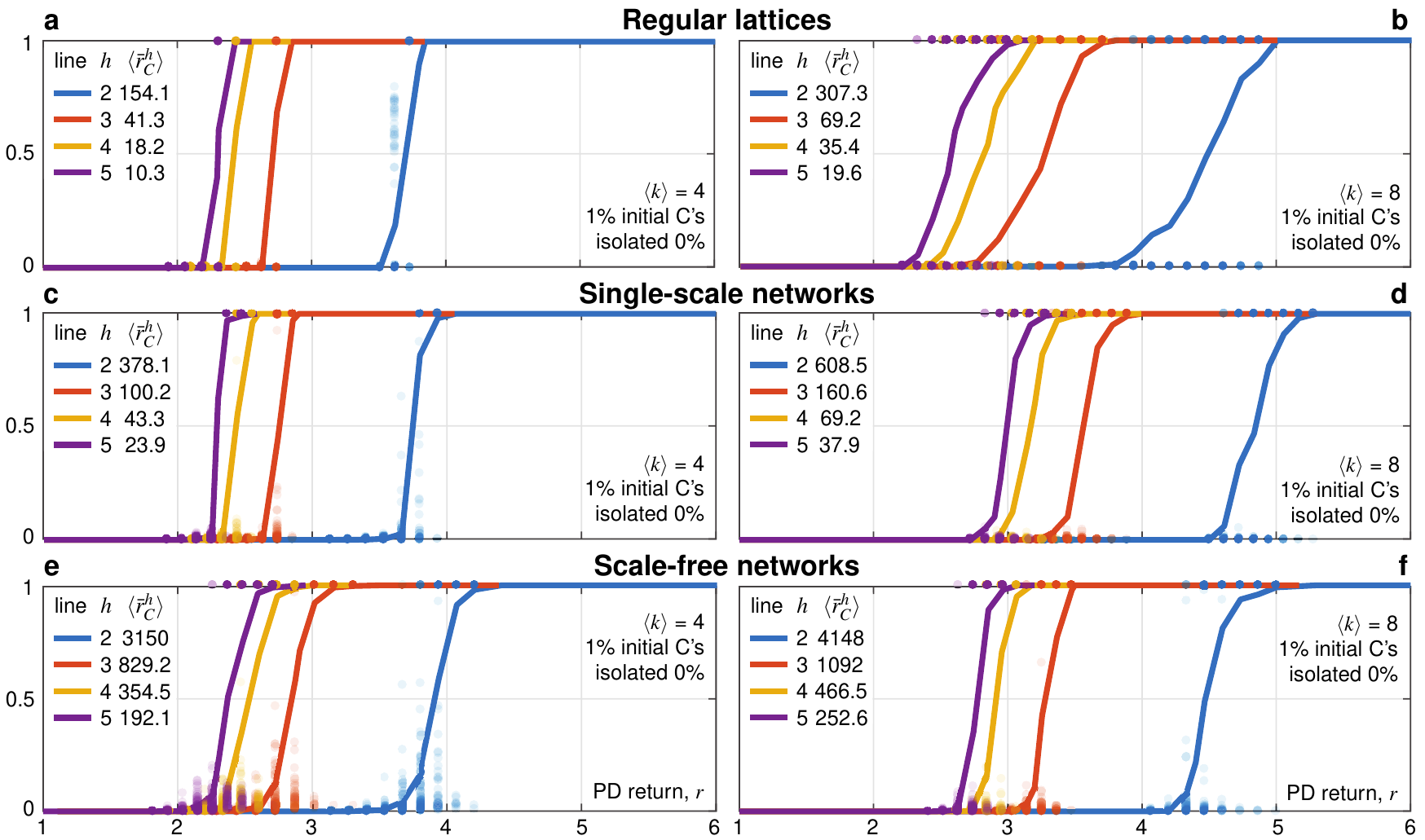}
\caption{Networked rational reciprocity under random pair placement of the initial C's.
The figure complements Fig.~\ref{fig:r01} by reproducing the simulations with random placement of the initial C's by randomly selecting pairs of C-neighbors instead of single individuals.
The $10$ initial C's ($1\%$ of $N=1000$ nodes) are iteratively selected as follows:
each time a selected node has no C-neighbors, the next node is selected among its neighbors;
the last node is either selected to pair the previous one, or among the D's connected to C's.
Both $r_{\mathrm{C},\min}^h$ and $r_{\mathrm{C},\max}^h$, as well as their gap, get reduced w.r.t. random placement in Fig.~\ref{fig:r01}.
Recall that isolated C's switch to D at first strategy revision, so that starting with all isolated C's it is always possible to end in the trivial stalemate all-D.
Note that degree-rank-pair-placement---selecting initial C's according to degree-ranking by pairing isolated ones---would have reproduced the results of Fig.~\ref{fig:r01}, as random and degree-rank selections are essentially equivalent for single-scale networks and degree-rank-C-placement is unlikely to leave isolated C's in scale-free networks.}
\label{fig:r2c}
\end{figure}


\begin{figure}[h]
\centering
\includegraphics{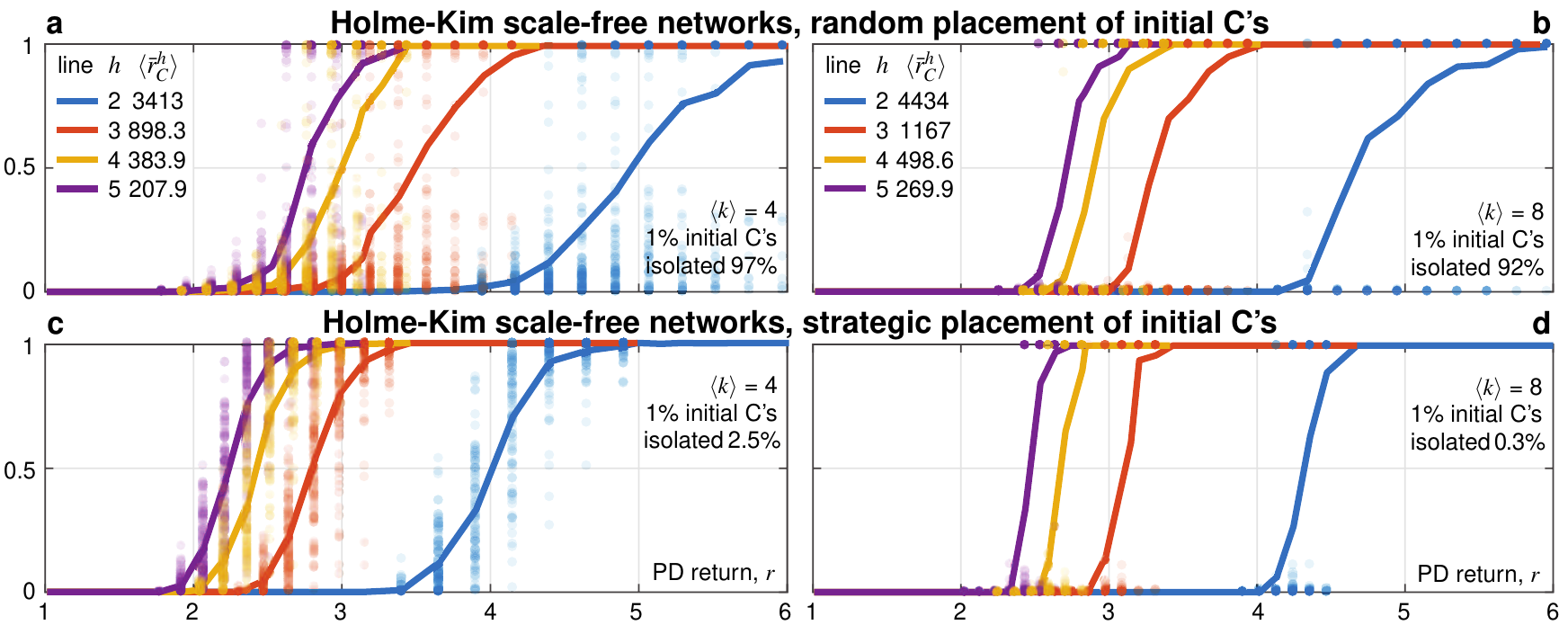}
\caption{Networked rational reciprocity in scale-free networks with medium-high transitivity.
We checked the robustness of our results with respect to a different type of scale-free network (Holme-Kim model,\cite{Holme_and_Kim_02_PRE} HK) with tunable transitivity (the average probability that the neighbors of a node are neighbors themselves, see SI note 4),
since network transitivity---shown to favor cooperation under imitation update\cite{Santos_et_al_06_PRSB,Assenza_et_al_08_PRE}---vanishes with the size of Barabási-Albert networks. \cite{Klemm_and_Eguiluz_02_PRE}
(See Sect.~\ref{sec:net} for details on the BA and HK algorithms.)
The resulting transitivity (average over $100$ networks) is $0.74$ for average degree $\langle k\rangle=4$ (left) and $0.28$ for $\langle k\rangle=8$ (right)
(respectively $0.027$ and $0.037$ in Fig.~\ref{fig:r01};
transitivity is known to increases/decrease with $\langle k\rangle$ in BA/HK networks \cite{Holme_and_Kim_02_PRE,Klemm_and_Eguiluz_02_PRE}).
Although the theoretical degree distribution for large size is the same of a BA network, the effect of raising transitivity in a finite network is an increase of nodes with low and high degree to the detriment of nodes with intermediate degree (checked on average on our $100$ networks).
This `finite size effect' explains our results:
slightly lower $r_{\mathrm{C},\min}^h$ and higher $r_{\mathrm{C},\max}^h$ compared with the same panel in Fig.~\ref{fig:r01}.
Invasion is indeed favored by the enhanced initial presence of low-connected D's, who change to C under mild returns if connected to a C, whereas fixation is hindered by the enhanced initial presence of D-hubs.
The increased number of simulations not ended in all-C or all-D
(dots in the open interval $(0,1)$, essentially in the left panels where transitivity is very high, w.r.t. Fig.~\ref{fig:r01})
can be also explained in terms of the loss of nodes with intermediate degree.
On one hand, the loss makes the evolution slower, possibly forming bottlenecks through which cooperation must percolate, so that some of the dots at significant C-levels might denote simulations ending in all-C on a longer timescale.
On the other hand, the loss makes nontrivial stalemates and fluctuations more frequent, as D-nodes with high degree can prevent the convergence to all-C even under degree-rank-C-placement.
(See Sect.~\ref{sec:num} for more details on the numerical results.)}
\label{fig:rcl}
\end{figure}

\clearpage

\section*{Supplementary Methods}

%
%
\subsection{The probability to play $p_t$}
\label{sec:pt}
After $t-1\ge 0$ consecutive abstentions following an exploitation by the neighbor $j$, the probability $p_{ij}$ that the C-individual $i$ agrees to play
with $j$ at the next game round is set to $p_t=1-(1-\delta_{\eps})^t$.
It is the probability that $j$ had revised
her strategy at least once ever since the exploitation, according to the reciprocity-biased rate of strategy update $\delta_{\eps}$.
(Recall that $\delta_{\eps}=(1-\eps)\delta\in(0,1)$ is equal to, resp.~smaller/larger than, the actual rate $\delta$ for for normally, resp.~super/sub, reciprocating individuals; $\eps=0$, resp.~$\eps\gtrless 0$.)

If the $(i,j)$ interaction takes place at the next round, then $i$ sets $p_{ij}$ to $1$ or to $p_1=\delta_{\eps}$ depending on whether $j$ cooperates or defects.
Otherwise, $i$ updates $p_{ij}$ to $p_{t+1}$.
Note that $p_{t+1}$ can be obtained with the recursion
\begin{equation}
\label{eq:ptr}
p_{t+1}=(1-\delta_{\eps})p_t+\delta_{\eps},\quad t\ge 1.
\end{equation}
We set $p_0=1$ and, whenever needed in the following, $t_{ij}$ denotes the integer giving $p_{t_{ij}}=p_{ij}$ for the neighbor pair $(i,j)$.

\subsection{The probability of getting exploited $P_{\mathrm{CD}}^{t}$}
\label{sec:pCDh}
Starting after a given game round with $p_{ij}=p_{t_{ij}}$ for some $t_{ij}\ge 0$
(or after initialization with $t_{ij}=0$),
$P_{\mathrm{CD}}^{t}(t_{ij})$ is the probability that the C-individual $i$ plays
with the D-neighbor $j$ at the $t$-th future round, assuming no change of strategy.
(The $t_{ij}$-argument, sometimes omitted in the following, makes initialization explicit.)
We therefore have $P_{\mathrm{CD}}^{1}(t_{ij})=p_{t_{ij}}$ for $t=1$, while for $t>1$ we use the recursion
\begin{equation}
\label{eq:pcdt}
P_{\mathrm{CD}}^{t+1}=P_{\mathrm{CD}}^{t} \delta_{\eps} + (1-P_{\mathrm{CD}}^{t})\big((1-\delta_{\eps})P_{\mathrm{CD}}^{t}+\delta_{\eps}\big)=
\delta_{\eps} + (1-\delta_{\eps})P_{\mathrm{CD}}^{t} (1-P_{\mathrm{CD}}^{t}).
\end{equation}
That is, if $i$ is exploited by $j$ at round $t$, she will then play at round $t+1$ with probability $p_1=\delta_{\eps}$
(first term after the first equal sign in~\eqref{eq:pcdt});
otherwise $P_{\mathrm{CD}}^{t}$ is updated as $p_t$ with the rule~\eqref{eq:ptr} (second term).

Starting with $p_{ij}=\delta_{\eps}$ (i.e., $t_{ij}=1$), the probabilities $p_t$ and $P_{\mathrm{CD}}^{t}$ are listed in Table~\ref{tab:pCDh} for $t\ge 1$ (first and second columns).
Both probabilities have linear ($1$-st-order) leading $\delta_{\eps}$-term with same coefficient $t$, i.e.,
$P_{\mathrm{CD}}^{t}(t_{ij})\approx (t_{ij}+t-1)\delta_{\eps}$ for small $\delta_{\eps}$
(up to $1$-st-order, only the terms $\delta_{\eps}$ and $P_{\mathrm{CD}}^{t}$ matter in the right-most side of~\eqref{eq:pcdt})

\begin{figure}[ht]
\centering
\includegraphics{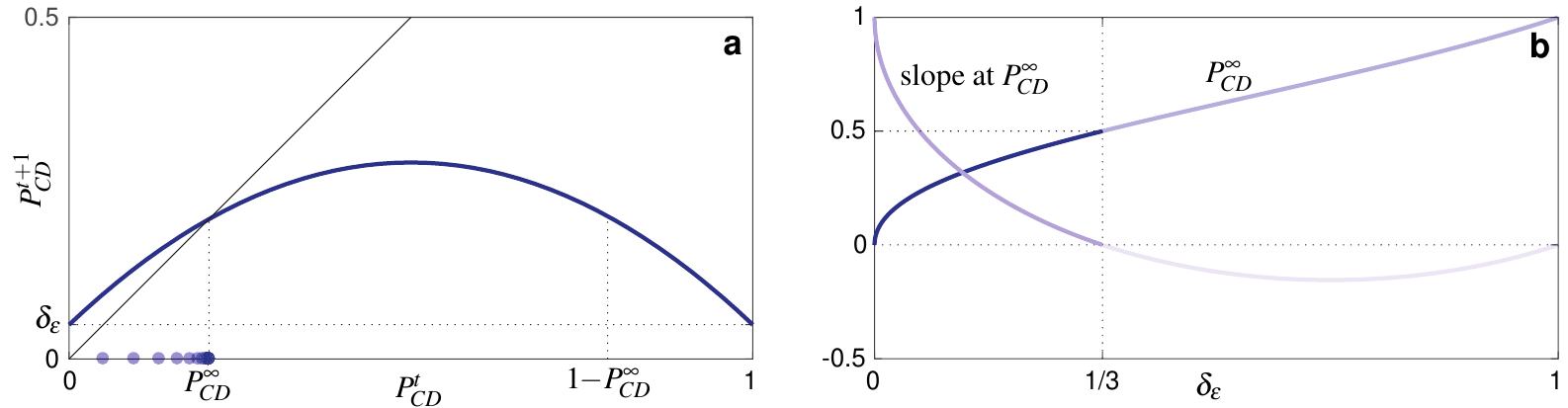}
\caption{The dynamics of the recursion \eqref{eq:pcdt}.
({\sffamily\textbf a})
The right-hand side of \eqref{eq:pcdt} for $\delta_{\eps}=0.05$ (solid thick line; the thin line is the diagonal)
and the probabilities $P_{\mathrm{CD}}^{t}(1)$, $t=1,\ldots,10$ (dots).
({\sffamily\textbf b})
The fixed point $P_{\mathrm{CD}}^{\infty}$ and the asymptotic rate of convergence 
(the local slope $1\!-\!\sqrt{\delta_{\eps}(4-3\delta_{\eps})}$ of the right-hand side of \eqref{eq:pcdt} at the fixed point)
as a function of $\delta_{\eps}$.
The slope is in $(-1,1)$ (asymptotic stability) for all $\delta_{\eps}\in(0,1)$
and in $(0,1)$ (monotonic convergence) for $\delta_{\eps}\in(0,1/3)$.}
\label{fig:pcdh}
\end{figure}

As $t\to\infty$, $P_{\mathrm{CD}}^{t}$ converges (independently of the initialization) to the infinite-horizon limit
\begin{equation}
\label{eq:pcdeq}
P_{\mathrm{CD}}^{\infty}=
\Frac{1}{2}\Frac{\sqrt{4\delta_{\eps}-3\delta_{\eps}^2}-\delta_{\eps}}{1-\delta_{\eps}}
\approx\sqrt{\delta_{\eps}}\;\;\text{for small $\delta_{\eps}$}.
\end{equation}
The limit is reached monotonically if $\delta_{\eps}<1/3$
(from below if $P_{\mathrm{CD}}^{1}<P_{\mathrm{CD}}^{\infty}$ or $P_{\mathrm{CD}}^{1}>1-P_{\mathrm{CD}}^{\infty}$;
from above if $P_{\mathrm{CD}}^{\infty}<P_{\mathrm{CD}}^{1}\le 1-P_{\mathrm{CD}}^{\infty}$;
see Fig.~\ref{fig:pcdh}).
This condition is met in the numerical analysis, where we use at most $\delta_{\eps}=0.1$
(corresponding to the perturbation of Fig.~\ref{fig:red} w.r.t. the baseline value $\delta_{\eps}=0.05$, see Table~\ref{tab:par}).

\subsection{The probability to reciprocate $P_{\mathrm{CC}}^{t}$}
\label{sec:pCCh}
Starting after a given game round with $p_{ij}=p_{t_{ij}}$ for some $t_{ij}\ge 0$
(or after initialization with $t_{ij}=0$),
$P_{\mathrm{CC}}^{t}(t_{ij})$ is the probability that the C-individual $i$ plays
at the $t$-th future round with the C-neighbor $j$, assuming no change of strategy.
If $t_{ij}=0$, then $i$ and $j$ reciprocated cooperation in the last round, so that $P_{\mathrm{CC}}^{t}(0)=1$ for all $t\ge 1$.
Otherwise, $i$ stopped playing after getting exploited by $j$ who turned C in the meanwhile (so that $p_{ji}=1$).
In both cases we have $P_{\mathrm{CC}}^{1}(t_{ij})=p_{t_{ij}}$ for $t=1$, while for $t>1$ we use the recursion
\begin{equation}
\label{eq:pcct}
P_{\mathrm{CC}}^{t+1}=P_{\mathrm{CC}}^{t} + (1-P_{\mathrm{CC}}^{t}) p_{t_{ij}+t}=
1-(1-\delta_{\eps})^{t_{ij}+t} + P_{\mathrm{CC}}^{t} (1-\delta_{\eps})^{t_{ij}+t}.
\end{equation}
That is, if $i$ does play at round $t$, she will certainly play at round $t+1$
(first term after the first equal sign in~\eqref{eq:pcct});
otherwise $P_{\mathrm{CC}}^{t}$ changes from $p_{t_{ij}+t-1}$ to $p_{t_{ij}+t}$ (second term).

Starting with $p_{ij}=\delta_{\eps}$ (i.e., $t_{ij}=1$), the probability $P_{\mathrm{CC}}^{t}$ is listed in Table~\ref{tab:pCDh} for $t\ge 1$ (third column).
The leading $\delta_{\eps}$-term is linear ($1$-st-order) and from \eqref{eq:pcct} it follows that its coefficient in $P_{\mathrm{CC}}^{t+1}$ is $t_{ij}+t$ plus the coefficient in $P_{\mathrm{CC}}^t$
(up to $1$-st-order, only the $1$-st-order term of the first power $(1-\delta_{\eps})^{t_{ij}+t}$ and the $0$-order term of the second matter in the right-most side of~\eqref{eq:pcct}).
For sufficiently small $\delta_{\eps}$, we hence have
\begin{equation}
\label{eq:pcccd}
P_{\mathrm{CC}}^{t}(t_{ij}) > P_{\mathrm{CD}}^{t}(t_{ij})\;\;\text{for any $t_{ij}\ge 0$ and $t>1$}.
\end{equation}
The probability $P_{\mathrm{CC}}^{t}$ monotonically increases to $1$ (independently of $t_{ij}$).

We now show that inequality \eqref{eq:pcccd} holds true for any $\delta_{\eps}\in(0,1)$.
From the first right-hand side in \eqref{eq:pcct} (time-shifted from $t-1$ to $t$), we see that $P_{\mathrm{CC}}^{t}(t_{ij})$ is an interior convex combination (by construction $0<P_{\mathrm{CC}}^{t-1}<1$) of $1$ and $p_{t_{ij}+t-1}<1$, so that $P_{\mathrm{CC}}^{t}(t_{ij}) > p_{t_{ij}+t-1}$ for any $t>1$.
Similarly, from \eqref{eq:pcdt} (time-shifted from $t-1$ to $t$), we see that $P_{\mathrm{CD}}^t(t_{ij})$ is an interior convex combination of $\delta_{\eps}$ (the smallest of the $p_t$ for $t\ge 1$) and the transformation of $P_{\mathrm{CD}}^{t-1}$ by rule~\eqref{eq:ptr}.
By construction, we hence have $P_{\mathrm{CD}}^t(t_{ij}) < p_{t_{ij}+t-1}$ for any $t>1$.

\begin{table}[t!]
\renewcommand{\arraystretch}{2.2}
\scalebox{.85}{
\begin{tabular}{llll}
\hline\\[-11.5mm]
\scalebox{1.16}{$t$} & \scalebox{1.16}{$p_t$} & \scalebox{1.16}{$P_{\mathrm{CD}}^{t}$} & \scalebox{1.16}{$P_{\mathrm{CC}}^{t}$}\\[-0.5mm]
\hline\\[-12mm]
$1$ & $\delta_{\eps}$ & $\delta_{\eps}$ & $\delta_{\eps}$\\[-4mm]
$2$ & $2 \delta_{\eps} - \delta_{\eps}^2$ &
$2 \delta_{\eps} - 2 \delta_{\eps}^2 + \delta_{\eps}^3$ &
$3 \delta_{\eps} - 3 \delta_{\eps}^2 + \delta_{\eps}^3$\\[-4mm]
$3$ & $3 \delta_{\eps} - 3 \delta_{\eps}^2 + \delta_{\eps}^3$ &
$3 \delta_{\eps} - 8 \delta_{\eps}^2 + 15 \delta_{\eps}^3 - 17 \delta_{\eps}^4 + 12 \delta_{\eps}^5
\cdots + \delta_{\eps}^7$ &
$6 \delta_{\eps} - 15 \delta_{\eps}^2 + 20 \delta_{\eps}^3 - 15 \delta_{\eps}^4
 + 6 \delta_{\eps}^5 - \delta_{\eps}^6$\\[-4mm]
$4$ & $4 \delta_{\eps} - 6 \delta_{\eps}^2 + 4 \delta_{\eps}^3 - \delta_{\eps}^4$ &
$4 \delta_{\eps} - 20 \delta_{\eps}^2 + 80 \delta_{\eps}^3 - 234 \delta_{\eps}^4 + 525 \delta_{\eps}^5
\cdots + \delta_{\eps}^{15}$ &
$10 \delta_{\eps} - 45 \delta_{\eps}^2 + 120 \delta_{\eps}^3 - 210 \delta_{\eps}^4 + 252 \delta_{\eps}^5
\cdots - \delta_{\eps}^{10}$\\[-4mm]
$5$ & $5 \delta_{\eps} - 10 \delta_{\eps}^2 + 10 \delta_{\eps}^3 - 5 \delta_{\eps}^4 + \delta_{\eps}^5$ &
$5 \delta_{\eps} - 40 \delta_{\eps}^2 + 276 \delta_{\eps}^3 - 1514 \delta_{\eps}^4 + 6871 \delta_{\eps}^5
\cdots + \delta_{\eps}^{31}$ &
$15 \delta_{\eps} - 105 \delta_{\eps}^2 + 455 \delta_{\eps}^3 - 1365 \delta_{\eps}^4 + 3003 \delta_{\eps}^5
\cdots + \delta_{\eps}^{15}$\\
\hline
\end{tabular}}
\caption{Probabilities $p_t$, $P_{\mathrm{CD}}^{t}(1)$, and $P_{\mathrm{CC}}^{t}(1)$
(i.e., starting with $P_{\mathrm{CD}}^{1}=p_1=\delta_{\eps}$).}
\label{tab:pCDh}
\end{table}

\subsection{The predicted payoff gain $\Delta\pi_{\mathrm{C}}^h$}
\label{sec:DpiC}
After any game round, each of the $k_i$ probabilities $p_{ij}$ of the C-individual $i$ is either $1$ or equal to $p_{t_{ij}}$ for some $t_{ij}\ge 1$.
In revising her strategy, $i$ assumes the strategy of $j$ unchanged since they last interacted.
That is, $i$ believes $j$ to be a C if they either both cooperated at the last round or $j$ did not play because the last time they interacted $i$ exploited $j$, so that $p_{ji}<1$, and then $i$ changed to C.
In both situations, $p_{ij}=1$.
Conversely, $i$ believes $j$ to be a D if the last time they interacted $j$ exploited $i$, so that $p_{ij}<1$.
C's can hence underestimate the number of their C-neighbors.
Let $k_{\mathrm{C},i}\,(\le k_i)$ be the number of C-neighbors of individual $i$ and $k'_{\mathrm{C},i}\,(\le k_{\mathrm{C},i})$ the number of $p_{ij}=1$.

Assuming no strategy change in the neighborhood within the predictive horizon $h$
(see Methods in the main text for a discussion of this assumption),
$i$ computes the expected payoffs accumulated in the next $h$ rounds behaving as D, $\pi_{\mathrm{CD}}^h$, or as C, $\pi_{\mathrm{CC}}^h$, as follows:
\begin{itemize}
\item
Initialize $\pi_{\mathrm{CD}}^h=0$ and $\pi_{\mathrm{CC}}^h=0$ and consider the sums
\vspace*{-3mm}
\begin{equation}
\label{eq:scd}
S^h_{\mathrm{CD}}(\tau)=\sum_{t=1}^h P_{\mathrm{CD}}^{t}(\tau)\quad\text{and}\quad
S^h_{\mathrm{CC}}(\tau)=\sum_{t=1}^h P_{\mathrm{CC}}^{t}(\tau)
\end{equation}

\vspace*{-3.5mm}\noindent
(note that $t_{ij}$ must be replaced with $\tau$ in the right-hand side of recursion~\eqref{eq:pcct} to define
$P_{\mathrm{CC}}^{t}(\tau)$ in terms of $P_{\mathrm{CC}}^{t}(\tau-\!1)$);
\item
For each $p_{ij}=1$, add $r\,S^h_{\mathrm{CD}}(t_{ji})$ to $\pi_{\mathrm{CD}}^h$;
\item
For each $p_{ij}=1$, add $(r-1)S^h_{\mathrm{CC}}(t_{ji})$ to $\pi_{\mathrm{CC}}^h$;
\item
For each $p_{ij}<1$, subtract $S^h_{\mathrm{CD}}(t_{ij})$ from $\pi_{\mathrm{CC}}^h$.
\end{itemize}
Then, $\Delta\pi_{\mathrm{C}}^h=\pi_{\mathrm{CD}}^h-\pi_{\mathrm{CC}}^h$, i.e.,
\begin{eqnarray}
\label{eq:dpiC}
\Delta\pi_{\mathrm{C}}^h &=&\!
\sum_{p_{ij}=1}\big(r S^h_{\mathrm{CD}}(t_{ji}) - (r-1) S^h_{\mathrm{CC}}(t_{ji})\big) + \sum_{p_{ij}<1} S^h_{\mathrm{CD}}(t_{ij}) \nonumber\\ &=& 
-r\!\sum_{p_{ij}=1}\big(
\underbrace{S^h_{\mathrm{CC}}(t_{ji})-S^h_{\mathrm{CD}}(t_{ji})}_{>0\;\text{for $h\ge 2$, see \eqref{eq:pcccd}}}\big)
+ \sum_{p_{ij}=1} S^h_{\mathrm{CC}}(t_{ji}) + \sum_{p_{ij}<1} S^h_{\mathrm{CD}}(t_{ij}).
\end{eqnarray}

Note that the indexes $t_{ji}$ are involved in the computation, i.e., if $j$ refused to play with $i$ at the last game round, we assume that $i$ remembers when she has last exploited $j$.
Also note that
\begin{equation}
\label{eq:dpiC1}
\Delta\pi_{\mathrm{C}}^1=\!
\sum_{p_{ij}=1} (r p_{ji} - (r-1) p_{ji}) + \sum_{p_{ij}<1} p_{ij} = 
\sum_{p_{ij}=1} p_{ji} + \sum_{p_{ij}<1} p_{ij} > 0,
\end{equation}
i.e., defect is the best-response for $h=1$.

\subsection{The predicted payoff gain $\Delta\pi_\mathrm{D}^h$}
\label{sec:DpiD}
After any game round, the D-individual $i$ has probability $p_{ij}=1$ toward each of her $k_i$ neighbors and full information, i.e., knowledge of the $p_{ji}$ of the $k_{\mathrm{C},i}\le k_i$ C-neighbors.
In revising her strategy
(assuming no strategy change in the neighborhood within the predictive horizon $h$, see Methods in the main text),
$i$ computes the expected payoffs accumulated in the next $h$ rounds behaving as C, $\pi_{\mathrm{DC}}^h$, or as D, $\pi_{\mathrm{DD}}^h$, as follows:
\begin{itemize}
\item
Initialize $\pi_{\mathrm{DC}}^h=0$ and $\pi_{\mathrm{DD}}^h=0$ and consider the sums $S^h_{\mathrm{CD}}$ and $S^h_{\mathrm{CC}}$ in~\eqref{eq:scd};
\item
For each C-neighbor $j$, add $(r-1)S^h_{\mathrm{CC}}(t_{ji})$ to $\pi_{\mathrm{DC}}^h$;
\item
For each C-neighbor $j$, add $rS^h_{\mathrm{CD}}(t_{ji})$ to $\pi_{\mathrm{DD}}^h$.
\item
For each D-neighbor $j$, subtract $S^h_{\mathrm{CD}}(1)$ from $\pi_{\mathrm{DC}}^h$;
\end{itemize}
Then, $\Delta\pi_\mathrm{D}^h=\pi_{\mathrm{DC}}^h-\pi_{\mathrm{DD}}^h$, i.e.,
\begin{eqnarray}
\label{eq:dpiD}
\Delta\pi_\mathrm{D}^h &=&\!
\sum_{m_j=\text{C}}\big((r-1) S^h_{\mathrm{CC}}(t_{ji}) - r S^h_{\mathrm{CD}}(t_{ji})\big) - \sum_{m_j=\text{D}} S^h_{\mathrm{CD}}(1) \nonumber\\ &=& 
r\!\sum_{m_j=\text{C}}\big(
\underbrace{S^h_{\mathrm{CC}}(t_{ji})-S^h_{\mathrm{CD}}(t_{ji})}_{>0\;\text{for $h\ge 2$, see \eqref{eq:pcccd}}}\big)
- \sum_{m_j=\text{C}} S^h_{\mathrm{CC}}(t_{ji}) - \sum_{m_j=\text{D}} S^h_{\mathrm{CD}}(1),
\end{eqnarray}
where $m_j\in\{\text{C},\text{D}\}$ is the strategy of individual $j$ and the sums span the neighborhood of individual $i$.

Note that
\begin{equation}
\label{eq:dpiD1}
\Delta\pi_\mathrm{D}^1=\!
\sum_{m_j=\text{C}} ((r-1) p_{ji}-r p_{ji}) - \sum_{m_j=\text{D}} p_1 = 
-\sum_{m_j=\text{C}} p_{ji} - (k-k_{\mathrm{C}}) p_1 < 0,
\end{equation}
i.e., defect is the best-response for $h=1$.

\subsection{Proof of condition~\eqref{eq:condrinf} and the threshold $r_{\mathrm{C}}^{\infty}$}
\label{sec:hinf}
With an infinite predictive horizon ($h\to\infty$), the sums $S^h_{\mathrm{CD}}(\tau)$ and $S^h_{\mathrm{CC}}(\tau)$ in~\eqref{eq:scd} diverge with $S^h_{\mathrm{CD}}/h\to P_{\mathrm{CD}}^{\infty}$ and $S^h_{\mathrm{CC}}/h\to 1$ independently of $\tau$.
Consequently, for a C- and a D-individual with degree $k$ and $k_{\mathrm{C}}\le k$ C-neighbors, 
the predicted payoff gains $\Delta\pi_{\mathrm{C}}^h$ and $\Delta\pi_\mathrm{D}^h$ 
(from eqs.~\eqref{eq:dpiC} and~\eqref{eq:dpiD}) are unbounded with
\begin{subequations}
\label{eq:dpiCDi}
\begin{eqnarray}
\label{eq:dpiCi}
\Delta\pi_{\mathrm{C}}^h/h & \to &
-rk'_{\mathrm{C}}(1-P_{\mathrm{CD}}^{\infty})+k'_{\mathrm{C}}+(k-k'_{\mathrm{C}})P_{\mathrm{CD}}^{\infty},\\
\label{eq:dpiDi}
\Delta\pi_\mathrm{D}^h/h & \to &
rk_{\mathrm{C}}(1-P_{\mathrm{CD}}^{\infty})-k_{\mathrm{C}}-(k-k_{\mathrm{C}})P_{\mathrm{CD}}^{\infty},
\end{eqnarray}
\end{subequations}
where, for a C, $k'_{\mathrm{C}}\le k_{\mathrm{C}}$ is the number of $p_{ij}=1$.

Solving $\Delta\pi_{\mathrm{C}}^h/h<0$ and $\Delta\pi_\mathrm{D}^h/h>0$ in the limits in~\eqref{eq:dpiCDi} gives condition~\eqref{eq:condrinf} in the main text.
The threshold $r_{\mathrm{C}}^{\infty}$ on $r$ above which cooperation fixates starting from any cluster of (at least two) C's is then obtained from~\eqref{eq:condrinf} by setting $k$ to the largest degree $k_{\max}$ in the network and $k_{\mathrm{C}}=1$, i.e.,
\begin{equation}
\label{eq:rCinf}
r_{\mathrm{C}}^{\infty} = 1+k_{\max} \Frac{P_{\mathrm{CD}}^{\infty}}{1-P_{\mathrm{CD}}^{\infty}}.
\end{equation}

\subsection{The effect of the predictive horizon $h$}
\label{sec:cond2}
From~eqs.~\eqref{eq:dpiC} and~\eqref{eq:dpiD}, it immediately follows that the contributions to the predicted payoff gains
$\Delta\pi_{\mathrm{C}}^h$ and $\Delta\pi_{\mathrm{D}}^h$
of one more prediction step are 
\begin{equation}
\label{eq:ddpiC}
\Delta\pi_{\mathrm{C}}^{h+1}-\Delta\pi_{\mathrm{C}}^h =
-r\!\sum_{p_{ij}=1}\big(
\underbrace{P_{\mathrm{CC}}^{h+1}(t_{ji})-P_{\mathrm{CD}}^{h+1}(t_{ji})}_{>0\;\text{for $h\ge 1$, see \eqref{eq:pcccd}}}\big)
+ \sum_{p_{ij}=1} P_{\mathrm{CC}}^{h+1}(t_{ji}) + \sum_{p_{ij}<1} P_{\mathrm{CD}}^{h+1}(t_{ij})
\end{equation}
and
\begin{equation}
\label{eq:ddpiD}
\Delta\pi_\mathrm{D}^{h+1}-\Delta\pi_\mathrm{D}^h =
r\!\sum_{m_j=\text{C}}\big(
\underbrace{P_{\mathrm{CC}}^{h+1}(t_{ji})-P_{\mathrm{CD}}^{h+1}(t_{ji})}_{>0\;\text{for $h\ge 1$, see \eqref{eq:pcccd}}}\big)
- \sum_{m_j=\text{C}} P_{\mathrm{CC}}^{h+1}(t_{ji}) - \sum_{m_j=\text{D}} P_{\mathrm{CD}}^{h+1}(1).
\end{equation}
For any $h\ge 1$, the first can be made arbitrarily negative, the second arbitrarily positive, by a sufficiently large $r$.
Any multi-step predictive horizon ($h\ge 2$) is hence sufficient for the evolution of rational reciprocity, provided the game return is large enough.

Instead, for a given $r$, the effect on cooperation of extending the predictive horizon is positive, i.e., 
$\Delta\pi_{\mathrm{C}}^h$ and $\Delta\pi_{\mathrm{D}}^h$ are respectively decreasing and increasing with $h\ge 1$, if
\begin{equation}
\label{eq:ddpiC0}
r>1+\Frac{\sum_{p_{ij}=1} P_{\mathrm{CD}}^{h+1}(t_{ji}) + \sum_{p_{ij}<1} P_{\mathrm{CD}}^{h+1}(t_{ij})}
{\sum_{p_{ij}=1}\big(P_{\mathrm{CC}}^{h+1}(t_{ji})-P_{\mathrm{CD}}^{h+1}(t_{ji})\big)}
\quad\text{and}\quad
r>1+\Frac{\sum_{m_j=\text{C}} P_{\mathrm{CD}}^{h+1}(t_{ji}) + \sum_{m_j=\text{D}} P_{\mathrm{CD}}^{h+1}(1)}
{\sum_{m_j=\text{C}}\big(P_{\mathrm{CC}}^{h+1}(t_{ji})-P_{\mathrm{CD}}^{h+1}(t_{ji})\big)},
\end{equation}
obtained by solving $\Delta\pi_{\mathrm{C}}^{h+1}-\Delta\pi_{\mathrm{C}}^h<0$ and $\Delta\pi_\mathrm{D}^{h+1}-\Delta\pi_\mathrm{D}^h>0$ for $r$ from~eqs.~\eqref{eq:ddpiC} and~\eqref{eq:ddpiD}.

\begin{figure}[t!]
\centering
\includegraphics{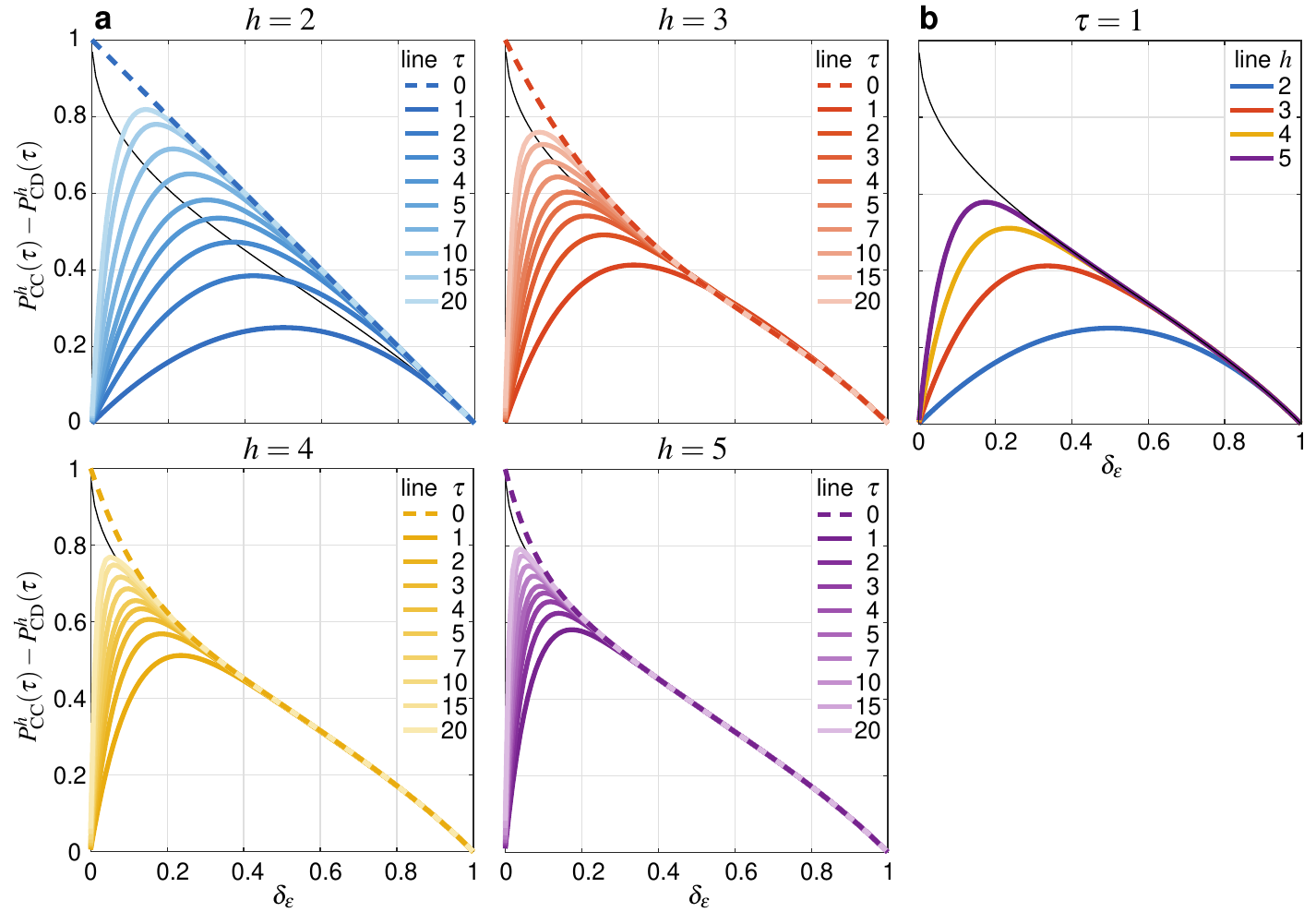}
\caption{Surplus of the probability to reciprocate $P_{\mathrm{CC}}^{h}(\tau)$ over the probability of getting exploited $P_{\mathrm{CD}}^{h}(\tau)$.
Plot of the surplus versus the reciprocity-biased rate of strategy update $\delta_{\eps}$.
({\sffamily\textbf a})
Different values of the predictive horizon ($h=2,\ldots,5$) and initializations of the probability to play at the next round ($\tau=0,\ldots,9$).
({\sffamily\textbf b})
The case $\tau=1$ in a single panel.
Colored lines: $P_{\mathrm{CC}}^{h}(\tau)-P_{\mathrm{CD}}^{h}(\tau)$;
black thin line: $1-P_{\mathrm{CD}}^{\infty}$ for comparison.
}
\label{fig:pccd}
\end{figure}

The denominators in \eqref{eq:ddpiC0} are minimized by $t_{ji}=1$.
Indeed, $t_{ji}=0$ gives $P_{\mathrm{CC}}^{h+1}(0)-P_{\mathrm{CD}}^{h+1}(0)=1-P_{\mathrm{CD}}^{h}(1)>1-P_{\mathrm{CD}}^{\infty}$
(note that we used $P_{\mathrm{CD}}^{h+1}(0)=P_{\mathrm{CD}}^{h}(1)$, though $P_{\mathrm{CD}}^{h+1}(\tau)\neq P_{\mathrm{CD}}^{h}(\tau+1)$ for $\tau\ge 1$).
And for $t_{ji}\ge 1$, $P_{\mathrm{CC}}^{h+1}(t_{ji})-P_{\mathrm{CD}}^{h+1}(t_{ji})$ increases with $t_{ji}$ for any $\delta_{\eps}\in(0,1)$ and remains below $1-P_{\mathrm{CD}}^{\infty}$ for $t_{ji}=1$
(checked numerically, see Fig.~\ref{fig:pccd}{\sffamily\textbf a} for $h=2,\ldots,5$).

Since the quantity $P_{\mathrm{CC}}^{h+1}(1)-P_{\mathrm{CD}}^{h+1}(1)$ increases with $h$ (and converges to $1-P_{\mathrm{CD}}^{\infty}$ for large $h$, see Fig.~\ref{fig:pccd}{\sffamily\textbf b}), a lower bound to the denominators in \eqref{eq:ddpiC0} is $P_{\mathrm{CC}}^2(1)-P_{\mathrm{CD}}^2(1)=\delta_{\eps}(1-\delta_{\eps})$ (see Table~\ref{tab:pCDh}) to be multiplied by $k'_{\mathrm{C}}$ (the number of $p_{ij}=1$) in the left condition and by $k_{\mathrm{C}}$ (the number of C-neighbors of individual $i$) in the right one.
An upper bound to the numerators is $k(1+3\delta_{\eps})/4$, obtained by replacing $P_{\mathrm{CD}}^{h+1}(t_{ji})$, $P_{\mathrm{CD}}^{h+1}(t_{ij})$, and $P_{\mathrm{CD}}^{h+1}(1)$ with the maximal value assumed by the right-most side of~\eqref{eq:pcdt}, i.e., with $\delta_{\eps}+\onefourth(1-\delta_{\eps})$.

We then have
$\Delta\pi_{\mathrm{C}}^{h+1}-\Delta\pi_{\mathrm{C}}^h<0$ and $\Delta\pi_\mathrm{D}^{h+1}-\Delta\pi_\mathrm{D}^h>0$ for any $h\ge 1$ under
\begin{equation}
\label{eq:condrinf2}
r>1+\Frac{k}{k_{\mathrm{C}}}\Frac{1+3\delta_{\eps}}{4\delta_{\eps}(1-\delta_{\eps})},
\end{equation}
Note that condition~\eqref{eq:condrinf2} might be very conservative, especially for small $\delta_{\eps}$.
Moreover, even if $\Delta\pi_{\mathrm{C}}^h$ and $\Delta\pi_\mathrm{D}^h$ are respectively increasing and decreasing with small $h$, they become and remain decreasing and increasing (under condition~\eqref{eq:condrinf} in the main text) for sufficiently large $h$.
Indeed, the right-hand sides in \eqref{eq:ddpiC0} converge to the right-hand side of~\eqref{eq:condrinf} for large $h$.

\subsection{The threshold $r_{\mathrm{C}}^h$}
\label{sec:rch}
With a finite horizon $h\ge 2$, it is already evident from eqs.~\eqref{eq:dpiC} and~\eqref{eq:dpiD} that, for any given network structure, there is a threshold $r_{\mathrm{C}}^h$ on $r$ above which
$\Delta\pi_{\mathrm{C}}^h<0$ and $\Delta\pi_\mathrm{D}^h>0$ for any C- and D-individual $i$
with at least one known C-neighbor (i.e., $p_{ij}=1$ if $i$ is a C, $m_j=\text{C}$ if $i$ is a D, for at least one $j$).
Solving $\Delta\pi_{\mathrm{C}}^h<0$ and $\Delta\pi_\mathrm{D}^h>0$ for $r$ from eqs.~\eqref{eq:dpiC} and~\eqref{eq:dpiD} gives
\begin{equation}
\label{eq:rCCDh}
r > 1+\Frac{\sum_{p_{ij}=1} S^h_{\mathrm{CD}}(t_{ji}) + \sum_{p_{ij}<1} S^h_{\mathrm{CD}}(t_{ij})}
{\sum_{p_{ij}=1}\big(S^h_{\mathrm{CC}}(t_{ji})-S^h_{\mathrm{CD}}(t_{ji})\big)}
\quad\text{and}\quad
r > 1+\Frac{\sum_{m_j=\text{C}} S^h_{\mathrm{CD}}(t_{ji}) + \sum_{m_j=\text{D}} S^h_{\mathrm{CD}}(1)}
{\sum_{m_j=\text{C}}\big(S^h_{\mathrm{CC}}(t_{ji})-S^h_{\mathrm{CD}}(t_{ji})\big)},
\end{equation}
from which we note that the $r$-threshold for a C to remain C (left) is typically lower than that for a D with same neighborhood to switch to C (right).
Indeed, the denominators in \eqref{eq:rCCDh} grow with $t_{ji}$, as each (positive) element of the sum
$S^h_{\mathrm{CC}}(t_{ji})-S^h_{\mathrm{CD}}(t_{ji})=\sum_{t=1}^h\big(P_{\mathrm{CC}}^{t}(t_{ji})-P_{\mathrm{CD}}^{t}(t_{ji})\big)$ does
(see Fig.~\ref{fig:pccd}{\sffamily\textbf a}), and the indexes $t_{ji}$ are expected to be higher for a C-individual $i$.

In the following we derive an upper bound to $r_{\mathrm{C}}^h$ consistent with the limit in \eqref{eq:rCinf}.
For any $h\ge 2$, the threshold $r_{\mathrm{C}}^h$ is the maximal value attained by the right-hand sides in \eqref{eq:rCCDh} over all possible choices of the node $i$ and over all possible configurations of its neighborhood
(in terms of the strategies $m_i$ and $m_j$ and of the probabilities $p_{ij}$ and $p_{ij}$), restricting however the search to configurations with at least one known C-neighbor that can be reached from an initial state (a state with $p_{ij}=p_{ji}=1$ for all connected pairs $(i,j)$ and $p_{ij}=p_{ji}=0$ otherwise).

\begin{figure}[t!]
\centering
\includegraphics{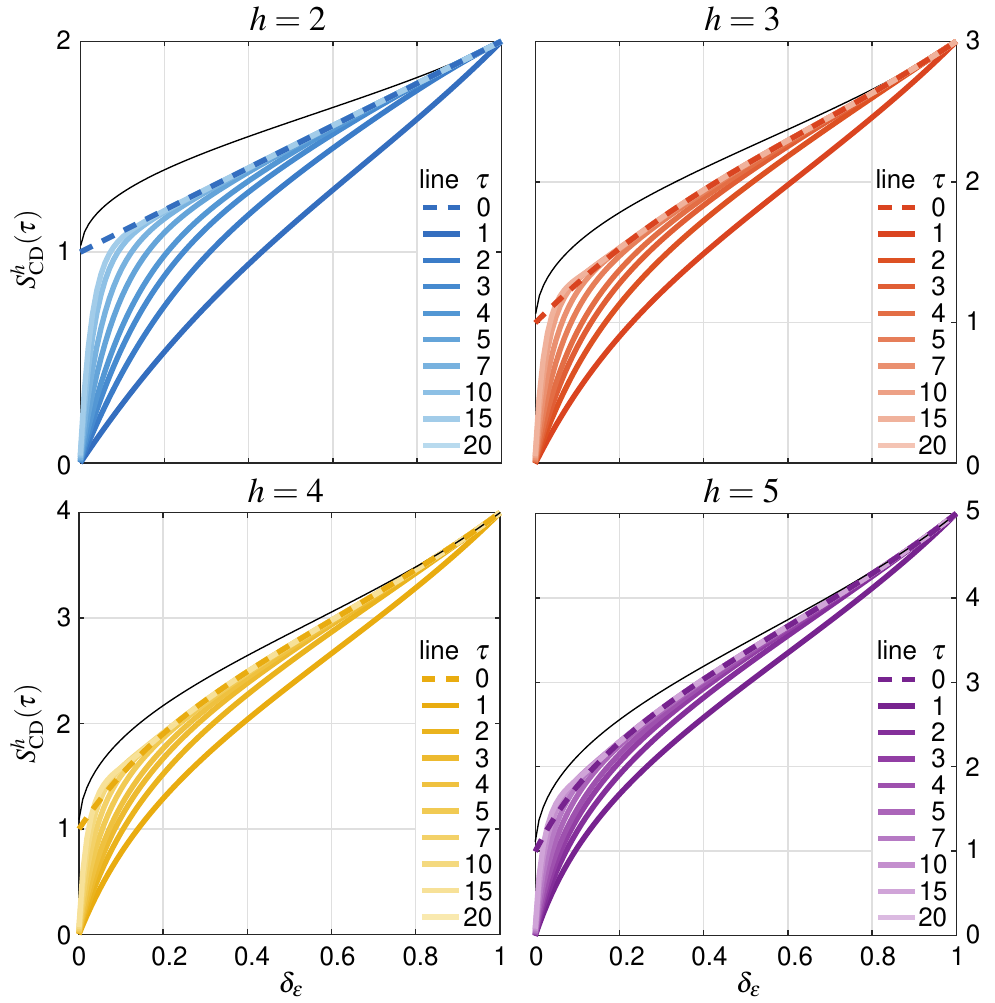}
\caption{The sum $S^h_{\mathrm{CD}}(\tau)$.
Plot of the sum versus the reciprocity-biased rate of strategy update $\delta_{\eps}$ for different values of the predictive horizon ($h=2,\ldots,5$) and initializations of the probability to play at the next round ($\tau=0,\ldots,9$).
Colored lines: $S^h_{\mathrm{CD}}(\tau)$;
black thin line: the upper bound $1+(h-1)P_{\mathrm{CD}}^{\infty}$.}
\label{fig:scd}
\end{figure}

As noted above, the denominators in \eqref{eq:rCCDh} are minimized by $t_{ji}=1$.
Moreover, the sum over $j$ comprises a single element if individual $i$ knows to have only one C-neighbor.
A lower bound to the denominators in \eqref{eq:rCCDh} is hence $S^h_{\mathrm{CC}}(1)-S^h_{\mathrm{CD}}(1)$, that can be easily be tabulated w.r.t. $h$ from Table~\ref{tab:pCDh}.
At numerator there are $k_i$ terms of the kind $S^h_{\mathrm{CD}}(\tau)$.
The value of $\tau\ge 0$ that maximizes $S^h_{\mathrm{CD}}(\tau)$ unfortunately depends on both $h$ and $\delta_{\eps}$, so we cannot upper bound the numerator by a specific choice of $\tau$.
However, we note that $S^h_{\mathrm{CD}}(\tau)$ is upper bounded by $1+(h-1)P_{\mathrm{CD}}^{\infty}$
(checked numerically, see Fig.~\ref{fig:scd} for $h=2,\ldots,5$).

The threshold $r_{\mathrm{C}}^h$ is hence upper bounded by
\begin{equation}
\label{eq:rChb}
r_{\mathrm{C}}^h < \bar r_{\mathrm{C}}^h =
1+k_{\max}\Frac{1+(h-1)P_{\mathrm{CD}}^{\infty}}{S^h_{\mathrm{CC}}(1)-S^h_{\mathrm{CD}}(1)} =
1+k_{\max}\Frac{\mbox{$\frac{\ss 1}{\ss h}$}+(1-\mbox{$\frac{\ss 1}{\ss h}$})P_{\mathrm{CD}}^{\infty}}
{\frac{S^h_{\mathrm{CC}}(1)}{h}-\frac{S^h_{\mathrm{CD}}(1)}{h}},
\end{equation}
obtained by taking $i$ as the node with highest degree $k_{\max}$.
The bound $\bar r_{\mathrm{C}}^h$ converges for large $h$ to $r_{\mathrm{C}}^{\infty}$
(from above under condition~\eqref{eq:condrinf};
$S^h_{\mathrm{CC}}(1)/h\to 1$ and $S^h_{\mathrm{CD}}(1)/h\to P_{\mathrm{CD}}^{\infty}$).
It is confirmed by all our simulations, though it can be quite conservative, especially for small $\delta_{\eps}$.

\subsection{Relevant probabilities for strategy update}
\label{sec:pup}
Two probabilities regarding the process of strategy update have been used in discussing the analytical results in the main text.
The first, $P'_{\text{update},k}$,
is the probability that, in a (possibly infinite) sequence of game rounds, one or more of the $k\ge 1$ neighbors of a given individual revise
their strategy before she does.
The second, $P''_{\text{update},k}$, 
is the probability that, in a (possibly infinite) sequence of game rounds in a star configuration, at least half of the $k\ge 2$ leaves revise
their strategy (at least once) before central one does
($k$ even).

To compute these two probabilities, recall that after each game round the probability that an individual with degree $k$ does not revise
strategy, while some of her neighbors do, is
\begin{equation}
\label{eq:pup3}
P'''_{\text{update},k} = (1-\delta)\big(1-(1-\delta)^k\big).
\end{equation}
Then, $P'_{\text{update},k}$ is obtained as the following geometric series
\begin{equation}
\label{eq:pup1}
P'_{\text{update},k}=
P'''_{\text{update},k}\sum_{t=0}^{\infty}(1-\delta)^{t(k+1)}=
P'''_{\text{update},k}\Frac{1}{1-(1-\delta)^{k+1}}=
1-\Frac{\delta}{1-(1-\delta)^{k+1}},
\end{equation}
where the $t$-th element of the sum is the probability that the strategy update occurs
after round $t+1$ (i.e., with no revision after the first $t$ rounds).
Note that, for small $\delta$, $P'_{\text{update},k}$ approaches $1-1/(k+1)$ from below
(use H{\^o}pital or note that $(1-\delta)^{k+1}\approx 1-(k+1)\delta+O(\delta^2)$ by the binomial expansion).

The probability $P''_{\text{update},k}$ is too complex to be characterized analytically.
However, for small $\delta$, we can assume that after each game round either none or at most one of the $k$ leaves of the star does revise strategy, i.e., we neglect the probability $\onehalf k(k-1)\delta^2(1-\delta)^{k-2}$ w.r.t. $k\delta(1-\delta)^{k-1}$.
Then, 
\begin{equation}
\label{eq:pup2}
P''_{\text{update},k}\,\approx\hspace{-1mm}
\prod_{k'=k/2+1}^{k}\!\!P'_{\text{update},k'}=
1-\Frac{1-(1-\delta)^{k/2}\hspace{2mm}}{1-(1-\delta)^{k+1}} =
1-\Frac{k/2}{k+1} + O(\delta) = \Frac{1}{2}+\Frac{1}{2(k+1)} + O(\delta).
\end{equation}
Thus, $P''_{\text{update},k}>\onehalf$ for small $\delta$, whereas it obviously drops to zero as $\delta\to 1$
(for $\delta=1$, all individuals revise strategy after each game round).
Numerically (with a Monte-Carlo approach), we have checked that for $\delta=0.05$ (the baseline value used in Fig.~\ref{fig:r01}) $P''_{\text{update},k}$ remains larger than $\onehalf$ for stars with up to $70$ nodes, that is not far from the maximal degree of the scale-free networks used in our simulations (see Table~\ref{tab:net}).

\subsection{Stalemate and fluctuations in the network of Fig.~\ref{fig:net}{\sffamily\textbf b}}
\label{sec:nf1}
Consider the network and the initial state of Fig.~\ref{fig:net}{\sffamily\textbf b},
with $i=1$, $j=2$, nodes $3$ to $k_1\hspace{-0.5mm}+\hspace{-0.3mm}1$ being the $k_1\hspace{-0.5mm}-\hspace{-0.3mm}1$ initial C-neighbors of $1$, and nodes $k_1\hspace{-0.5mm}+\hspace{-0.3mm}2$ to $k_1\hspace{-0.5mm}+\hspace{-0.3mm}k_2$ being the $k_2-1$ initial D-neighbors of $2$, $k_1,k_2>2$.

After the first game round, at which all individuals participate (i.e., no C abstains from playing), the strategy-revising C-individual $1$ remains C if
\begin{equation}
\label{eq:rC0}
\hspace{0mm}
r > r_{\mathrm{C},1}^0 =
1\hspace{-0.3mm}+\Frac{(k_1\hspace{-0.5mm}-\hspace{-0.3mm}1)S^h_{\mathrm{CD}}(0) + S^h_{\mathrm{CD}}(1)}
{(k_1\hspace{-0.5mm}-\hspace{-0.3mm}1)(h-S^h_{\mathrm{CD}}(0))} =
1\hspace{-0.3mm}+\Frac{S^h_{\mathrm{CD}}(0)}{h-S^h_{\mathrm{CD}}(0)} +
\Frac{1}{k_1\hspace{-0.5mm}-\hspace{-0.3mm}1}\Frac{S^h_{\mathrm{CD}}(1)}{h-S^h_{\mathrm{CD}}(0)}
\end{equation}
(from the left inequality in~\eqref{eq:rCCDh}, taking into account that $1$ lowered to $p_1$ the probability to play with $2$ at second round, i.e., $t_{12}=1$).
The strategy-revising D-individual $2$ changes to C if
\begin{equation}
\label{eq:rD0}
\hspace{0mm}
r > r_{\mathrm{D},2}^0 =
1\hspace{-0.3mm}+k_2\Frac{S^h_{\mathrm{CD}}(1)}
{S^h_{\mathrm{CC}}(1)-S^h_{\mathrm{CD}}(1)}
\end{equation}
(from the right inequality in~\eqref{eq:rCCDh}).
Strategy-revising C-individuals $3$ to $k_1\hspace{-0.5mm}+\hspace{-0.3mm}1$ remain C under
\begin{equation}
\label{eq:rC3}
\hspace{0mm}
r > r_{\mathrm{C},3}^0 =
1\hspace{-0.3mm}+\Frac{S^h_{\mathrm{CD}}(0)}{h-S^h_{\mathrm{CD}}(0)},
\end{equation}
(from the left inequality in~\eqref{eq:rCCDh} with $t_{i1}=t_{1i}=0$, $i=3,\ldots,k_1\hspace{-0.5mm}+\hspace{-0.3mm}1$), a condition implied by \eqref{eq:rC0}.
Strategy-revising D-individuals $k_1\hspace{-0.5mm}+\hspace{-0.3mm}2$ to $k_1\hspace{-0.5mm}+\hspace{-0.3mm}k_2$ remain D because have no C-neighbors.

Assume that no one changes.
After the second game round, if $1$ played with $2$, the situation is the same as after the first round;
otherwise, after $a$ consecutive abstentions, the strategy-revising C-individual $1$ remains C if
\begin{equation}
\label{eq:rCa}
r > r_{\mathrm{C},1}^{a} =
1\hspace{-0.3mm}+\Frac{(k_1\hspace{-0.5mm}-\hspace{-0.3mm}1)S^h_{\mathrm{CD}}(0) + S^h_{\mathrm{CD}}(a\hspace{-0.2mm}+\hspace{-0.3mm}1)}
{(k_1\hspace{-0.5mm}-\hspace{-0.3mm}1)(h-S^h_{\mathrm{CD}}(0))} =
1\hspace{-0.3mm}+\Frac{S^h_{\mathrm{CD}}(0)}{h-S^h_{\mathrm{CD}}(0)} +
\Frac{1}{k_1\hspace{-0.5mm}-\hspace{-0.3mm}1}\Frac{S^h_{\mathrm{CD}}(a\hspace{-0.2mm}+\hspace{-0.3mm}1)}{h-S^h_{\mathrm{CD}}(0)},
\end{equation}
while the strategy-revising D-individual $2$ changes to C if
\begin{equation}
\label{eq:rDa}
r > r_{\mathrm{D},2}^{a} =
1\hspace{-0.3mm}+\Frac{S^h_{\mathrm{CD}}(a\hspace{-0.2mm}+\hspace{-0.3mm}1) + (k_2\hspace{-0.5mm}-\hspace{-0.3mm}1)S^h_{\mathrm{CD}}(1)}
{S^h_{\mathrm{CC}}(a\hspace{-0.2mm}+\hspace{-0.3mm}1)-S^h_{\mathrm{CD}}(a\hspace{-0.2mm}+\hspace{-0.3mm}1)}.
\end{equation}
Again, strategy-revising C-individuals $3$ to $k_1\hspace{-0.5mm}+\hspace{-0.3mm}1$ remain C under \eqref{eq:rC3} and
strategy-revising D-individuals $k_1\hspace{-0.5mm}+\hspace{-0.3mm}2$ to $k_1\hspace{-0.5mm}+\hspace{-0.3mm}k_2$ do not change strategy.

The threshold $r_{\mathrm{C},1}^{a}$ is larger than $r_{\mathrm{C},1}^{0}$ for small $a$
($S^h_{\mathrm{CD}}(a\hspace{-0.2mm}+\hspace{-0.3mm}1)$ grows with $a\ge 0$ as long $a$ is sufficiently small, see Fig.~\ref{fig:scd})
and decreases with $k_1$, while $r_{\mathrm{D},2}^{a}$ decreases with $a$ for sufficiently large $k_2$
($S^h_{\mathrm{CC}}(a\hspace{-0.2mm}+\hspace{-0.3mm}1)-S^h_{\mathrm{CD}}(a\hspace{-0.2mm}+\hspace{-0.3mm}1)$ grows with $a\ge 0$ as discussed in Sect.~\ref{sec:rch})
and increases with $k_2$.
For sufficiently large $k_1$ and $k_2$, we therefore have $\max_a r_{\mathrm{C},1}^{a} < \min_a r_{\mathrm{D},2}^{a}$.
However, the opposite relation is also possible for any $a,k_1,k_2$, provided $\delta_{\eps}$ is sufficiently small.

If $\max_a r_{\mathrm{C},1}^{a}<r<\min_a r_{\mathrm{D},2}^{a}$, the network is in a stalemate.
The network can also reach a stalemate. Imagine node $2$ to be initially a C, with its condition to remain such after the first round unsatisfied, i.e.,
\begin{equation}
\label{eq:s1}
r<1\hspace{-0.3mm}+\Frac{S^h_{\mathrm{CD}}(0)+(k_2-1)S^h_{\mathrm{CD}}(1)}{h-S^h_{\mathrm{CD}}(0)},
\end{equation}
(possible for sufficiently large $k_2$)
and the condition for $1$ to remain C met, that is $r>r_{\mathrm{C},3}^0$ from \eqref{eq:rC3}.
Then, if $2$ changes to D the network goes in the stalemate.

But the same network can also produce long-term fluctuations.
Consider the initial state of Fig.~\ref{fig:net}{\sffamily\textbf b} with $r_{\mathrm{C},1}^0<r<r_{\mathrm{C},1}^{a}<\min_a r_{\mathrm{D},2}^{a}$ for some $a>0$.
Then, the C-individual $1$ is the only willing to change after $a$ consecutive abstentions.
Once $1$ switches to D, she will remain such for some game rounds.
If in the meantime the C-neighbors $3$ to $k_1\hspace{-0.5mm}+\hspace{-0.3mm}1$ do not change strategy and abstain for $a$ rounds, the condition for $1$ to go back C becomes
\begin{equation}
\label{eq:fl2}
r>1\hspace{-0.3mm}+\Frac{(k_1-\hspace{-0.3mm}1)S^h_{\mathrm{CD}}(a\hspace{-0.2mm}+\hspace{-0.3mm}1)+S^h_{\mathrm{CD}}(1)}
{(k_1-\hspace{-0.3mm}1)(S^h_{\mathrm{CC}}(a\hspace{-0.2mm}+\hspace{-0.3mm}1)-S^h_{\mathrm{CD}}(a\hspace{-0.2mm}+\hspace{-0.3mm}1))},
\end{equation}
that is satisfied for large enough $a$
($S^h_{\mathrm{CD}}(a\hspace{-0.2mm}+\hspace{-0.3mm}1)$ converges to $S^h_{\mathrm{CD}}(0)$ for large $a$, so that the right-hand side of \eqref{eq:fl2} approaches $r_{\mathrm{C},1}^0$).
Once $1$ switches back to C after a sufficiently long abstention of nodes $3$ to $k_1\hspace{-0.5mm}+\hspace{-0.3mm}1$, the situation is essentially the one just after the first game round.

The above example shows that long-term fluctuations are possible, though sometime difficult to observe. They might require specific sequences of events.
Essentially, once an individual changes strategy and plays a game round, she is not in the condition to switch back.
A D changes to C when the probabilities that most of her C-neighbors play in the next rounds are sufficiently high, otherwise she will mostly play with D-neighbors; and the strategy change further raises such probabilities, making the switch to D unattractive in the near future.
As well, switching to D lowers the probabilities that the C-neighbors play in the next rounds, thus preventing the near switch to C.

\subsection{Networks}
\label{sec:net}
We used six standard types of networks---three regular and three random---of $N=1000$ nodes and $M=N\langle k\rangle/2$ links, $\langle k\rangle$ denoting the average degree.
For each random type, we generated $100$ networks.
Details on the networks' structure and generation algorithms are given below.
Table~\ref{tab:net} reports several structural indicators (averaged over the $100$ generated networks for random types).
See Ref.~\onlinecite{BoLa:06} for further details on generation and analysis of complex networks.

\begin{table}[t!]
\setlength\tabcolsep{2mm}
\centering
\scalebox{.85}{
\begin{tabular}{lccccccc}
\hline\\[-3.5mm]
\multirow{2}{*}{\scalebox{1.16}{Network}} & 
\multirow{2}{*}{\scalebox{1.16}{$\langle k\rangle$}} &
\multirow{2}{*}{\scalebox{1.16}{$\sigma_k$}} &
\multirow{2}{*}{\scalebox{1.16}{$k_{\min}$}} &
\multirow{2}{*}{\scalebox{1.16}{$k_{\max}$}} &
\multirow{2}{*}{\scalebox{1.16}{Diameter}} & 
\scalebox{1.16}{Average} & 
\scalebox{1.16}{Transitivity}\\
 & & & & & & 
\scalebox{1.16}{distance} &
\scalebox{1.16}{}\\
\hline\\[-3mm]
Planar lattice & 4 & 0 & 4 & 4 & 32 & 16.3 & 0                \\
                & 8 & 0 & 8 & 8 & 20 & 11.3 & 0.42            \\[1mm]
Ring lattice   & 4 & 0 & 4 & 4 & 250&125.38 & 0.5             \\[1mm]
Complete network&999& 0 &999&999& 1 & 1 & 1                   \\[1mm]
Watts-Strogatz  & 4 & 1.40 & 2 & 9.85 & 9.0 & 5.32 & 0.003    \\
                & 8 & 1.99 & 4 & 15.87 & 5.3 & 3.59 & 0.007   \\[1mm]
Barabási-Albert & 4 & 5.24 & 2 & 82.27 & 7.32 & 4.07 & 0.027  \\
                & 8 & 8.81 & 4 & 108.6 & 5.0 & 3.17 & 0.037   \\[1mm]
Holme-Kim       & 4 & 5.40 & 2 & 89.13 & 11.06 & 4.88 & 0.738 \\
                & 8 & 8.96 & 4 & 115.8 & 5.44 & 3.26 & 0.284  \\[0.5mm]
\hline
\end{tabular}}
\caption{Networks' structural indicators (averaged over $100$ networks for random models).
Columns $\langle k\rangle$, $\sigma_k$, $k_{\min}$, and $k_{\max}$ respectively report the average, standard deviation, min and max of the nodes' degree.
Network size $N=1000$ nodes; number of links $M=N\langle k\rangle/2$.}
\label{tab:net}
\end{table}

\subsubsection{Regular networks}

\vspace*{1mm}\noindent{\it Planar lattices}: rectangular lattices of $N$ nodes with degree $k=4$ (horizontal and vertical links---square lattices) and $k=8$ (also including diagonal links) with periodic boundary conditions.
We used lattices of $40\times 25$ nodes.

\vspace*{1mm}\noindent{\it Ring lattices}: loops of $N$ nodes each connected to the $k/2$ nearest nodes in both left and right directions in the loop.
We used only $k=4$ (the degree $k$ must be an even integer).

\vspace*{1mm}\noindent{\it Complete network}: each node is connected to all $N-1$ others.

\subsubsection{Random networks}

\vspace*{1mm}\noindent{\it Watts-Strogatz (WS) with full rewiring}:
WS rewiring of all left links of a degree-$k$ ring lattice \citeSI{Watts98}.
We use this model to generate single-scale random networks, i.e., networks with sufficiently narrow degree distribution---small variance---so that the average degree $\langle k\rangle=k$ well describes the `scale' of the connections.
The standard single-scale model is the Erd\"os–Rényi\citeSI{BoLa:06} (ER) random network, where each of the $M=N\langle k\rangle/2$ links is included with probability
$p=M/(N(N-1)/2)=\langle k\rangle/(N-1)$
and the binomial degree distribution---$\mathrm{binomial}(k,N-1,p)$---converges for large $N$ to the Poisson with parameter $\langle k\rangle$.
The resulting network is however disconnected if $\langle k\rangle/N$ is too small
(it is disconnected with probability $1$ if $p<\ln N/N$, a condition that is met in our simulations with $N=1000$ and $\langle k\rangle=4$).
Though connectivity can be easily forced, the effects on the degree distribution are not easily quantifiable.
We therefore opted for the WS model that grants (almost sure) connectivity by the unaltered right-links of the initial ring lattice.
Note that the WS model is typically used with low rewiring probability (of the left links) to show the `small-world' property
(significant network transitivity and small diameter), whereas we use it here with full rewiring to better approximate an ER network.
Indeed, the resulting degree distribution\citeSI{barrat2000properties},
\begin{equation}
\label{eq:WS}
P(k)=\left\{\begin{array}{l}0\quad\text{if}\;k<\langle k\rangle/2,\;\text{otherwise}\\
\mathrm{binomial}(k-\langle k\rangle/2,(N-1)\langle k\rangle/2,1/(N-1))\end{array}\right.
\xrightarrow[N\hspace{-0.1mm}\to\infty]{}
\left\{\begin{array}{l}0\quad\text{if}\;k<\langle k\rangle/2,\;\text{otherwise}\\
(\langle k\rangle/2)^{k-\langle k\rangle/2}/(k-\langle k\rangle/2)!\exp(-\langle k\rangle/2)\end{array}\right.\hspace{-1.0mm},
\end{equation}
is Poissonian-like for large $N$
(see, e.g., Fig.~\ref{fig:net}{\sffamily\textbf b}).

\vspace*{1mm}\noindent{\it Barabási-Albert (BA)}:
BA degree-rank preferential attachment of $\langle k\rangle/2$ links. \citeSI{Albert_and_Barabasi_99_SCI}
We use this model to generate scale-free random networks, i.e., networks with broad degree distribution---large variance---showing low- and high-connected nodes, in spite of their average degree.
The BA algorithm produces networks with degree distribution that is zero for $k<\langle k\rangle/2$ and converges, for large $k$ and $N$, to a power law with exponent $-3$, hence showing an large variance (infinite variance in the limit $N\to\infty$).
The transitivity---the average over the network's nodes of the fraction of connected neighbor pairs---increases with $\langle k\rangle$ but vanishes as $(\ln N)^2/N$ for large $N$. \citeSI{Klemm_and_Eguiluz_02_PRE}

\vspace*{1mm}\noindent{\it Holme-Kim (HK)}:
HK scale-free networks with tunable transitivity \citeSI{Holme_and_Kim_02_PRE}.
We use this model to generate scale-free random networks with non-vanishing transitivity for large size.
With a tunable probability (that we set to $1$), the HK algorithm alternates steps of degree-rank preferential attachment
with steps in which a triangle is closed between the new node, the last preferred node and a neighbor of the latter.
The resulting transitivity does not vanish with $N$ and decreases with $\langle k\rangle$, because the higher number of triangles closed with larger $\langle k\rangle$ does not compensate for the increased number of possibilities.
The theoretical degree distribution for large $N$ is the same of the BA model.
However, w.r.t. a finite BA network, the HK algorithm raises the number of low- and high-connected nodes to the detriment of nodes with intermediate degree
(checked on average in our $100$ networks).
Imagine, e.g., to raise transitivity by rewiring. One option is to increase the number of hub-hub connections, to close triangles with common leaves.
This is achieved by detaching the termination of a hub's link to reconnect it to another hub. The latter node gains a link, while the looser moves left in the degree distribution.
For another option to increase transitivity in scale-free networks---by introducing the small-world property---see Ref.~\onlinecite{Klemm_and_Eguiluz_02_PRE}.

\subsection{Numerical simulations}
\label{sec:num}
For each panel of Figs.~\ref{fig:r01} and~\ref{fig:r50}--\ref{fig:rcl} and for each value of the predictive horizon ($h=2,\ldots,5$), we have first run simulations to identify the thresholds $r_{\mathrm{C},\min}^h$ and $r_{\mathrm{C},\max}^h$ (all simulations reach all-D for $r\le r_{\mathrm{C},\min}^h$; all-C for $r\ge r_{\mathrm{C},\max}^h$). 
We have then run simulations for $r$ in the open interval $(r_{\mathrm{C},\min}^h,r_{\mathrm{C},\max}^h)$ using an equally spaced grid with resolution of about $0.2$ (except for Fig.~\ref{fig:rcn}{\sffamily\textbf b} in which the scale of $r$ is larger).
For a given network of $N$ nodes (labeled $1$ to $N$) and assigned model parameters ($\delta$, $\eps$, $h$, $r$) (see Table~\ref{tab:par}), we used the following simulation procedure
(implemented in Matlab).

\begin{table}[t!]
\renewcommand{\arraystretch}{2.2}
\centering
\scalebox{.85}{
\begin{tabular}{clcc}
\hline\\[-11mm]
\scalebox{1.16}{Parameter} & \scalebox{1.16}{Description} & \scalebox{1.16}{Baseline values} & \scalebox{1.16}{Other values} \\[-0.5mm]
\hline\\[-11mm]
$N$      & Network size            & $1000$             & -- \\[-4mm]
$\delta$ & rate of strategy update & $0.05$             & $0.025$, $0.1$ \\[-4mm]
$\eps$   & reciprocity             & $0$                & $0.5$, $-1$    \\[-4mm]
$\delta_{\eps}$ & reciprocity-biased $\delta$ & $0.05$ & $(1-\eps)\delta$ \\[-4mm]
$h$      & predictive horizon      & $2$, $3$, $4$, $5$ & --             \\[-4mm]
$r$      & PD game return          & $[1, 6]$ & $[20, 60]$ in Fig.~\ref{fig:rcn}{\sffamily\textbf b}\\
\hline
\end{tabular}}
\caption{Model parameters.}
\label{tab:par}
\end{table}

\subsubsection*{Initial state}

\vspace*{1mm}\noindent{\it Initial C-level}:
we used either $1\%$ or $50\%$.
Let $N_{\mathrm{C}}$ denote the number of initial C's, i.e., $N_{\mathrm{C}}=10$ or $500$ for our networks of $N=1000$ nodes. 

\vspace*{1mm}\noindent{\it Random placement of the initial C's}:
The first $N_{\mathrm{C}}$ nodes of a random (uniform distribution) permutation of the integers $\{1,\ldots,N\}$ are set to C;
all others are set to D.

\vspace*{1mm}\noindent{\it Degree-rank placement of the initial C's}:
the $N_{\mathrm{C}}$ nodes with highest degree are set to C
(random choice, if needed, among the nodes sharing the smallest selected degree);
all others are set to D.

\vspace*{1mm}\noindent{\it Probabilities to play}:
$p_{ij}=p_{ji}=1$ for all connected pairs $(i,j)$; $p_{ij}=p_{ji}=0$ otherwise.

\subsubsection*{Random number generation}
For each simulation, we used two independent random number generators, one to generate the network (for random network models) and the initial state,
and one to perform the simulation (game interaction and strategy update).
The pair of initialization seeds for the two generators uniquely identify the simulation and are stored to allow reproduction.

\subsubsection*{Simulation length and outcome}

\vspace*{1mm}\noindent{\it Max. number of game rounds}:
$500/\delta$ game rounds ($=10^4$ for our baseline value of $\delta$).
It is the time length within which each individual revises strategy $500$ times, on average.

\vspace*{1mm}\noindent{\it Early termination}:
termination in all-C or all-D before the last game round.
There is no reason to continue the simulation, as all-C and all-D are both invariant states (trivial stalemates).

\vspace*{1mm}\noindent{\it Outcome}: $1$/$0$ in case of termination in all-C/D; otherwise,
the average C-level (fraction of C-nodes) over the last $100/\delta$ ($20\%$ of) game rounds (after strategy update).

\subsubsection*{Classification}

\vspace*{1mm}\noindent{\it Trivial stalemate}:
termination in all-C or all-D.

\vspace*{1mm}\noindent{\it Nontrivial stalemate}:
termination different from all-C and all-D with no strategy change in the last $100/\delta$ ($20\%$ of) game rounds.
There is of course no guarantee that this criterion identifies real stalemates. 
In practice, we classify as nontrivial stalemates the cases in which evolution slows down so much to be in a `practical' stalemate.

\vspace*{1mm}\noindent{\it Long-term fluctuation}:
termination different from all-C and all-D with some strategy change in the last $100/\delta$ game rounds yielding a regression slope within $10^{-6}$ (the slope of the regression line over the last $100/\delta$ game rounds does not exceed $\pm 1$ individual over $1000$ game rounds).
This is also a practical criterion.

\vspace*{1mm}\noindent{\it Non-convergence}:
termination different from all-C and all-D with some strategy change in the last $100/\delta$ game rounds yielding a regression slope exceeding $\pm 10^{-6}$.

\subsubsection*{Results}
We applied the above classifications to all simulations of Figs.~\ref{fig:r01} and~\ref{fig:r50}--\ref{fig:rcl} performed for $r_{\mathrm{C},\min}^h<r<r_{\mathrm{C},\max}^h$, i.e., for values of the game return $r$ and the predictive horizon $h$ for which neither all simulations ended in all-C nor in all-D.
The results, grouped by type of network and initialization (i.e., by figure panel) are reported in Table~\ref{tab:sim}.
(W.r.t. the predictive horizon $h$ within each panel, we note that the occurrence of stalemates and fluctuations and the corresponding levels of C slightly increase up to $h=4$ or $5$).
Only for Fig.~\ref{fig:r01}, the simulations not ended in all-C or all-D have been extended to a ten-times longer timescale to validate the classification
(see rows labeled `l.t.' in Table~\ref{tab:sim}).

\vspace*{2mm}\noindent{\it Trivial stalemates}\\[-6mm]
\begin{itemize}
\item
They constitute a significant fraction of outcomes independently of the network's structure and initialization, all-D dominating close to $r_{\mathrm{C},\min}^h$, all-C close to $r_{\mathrm{C},\max}^h$.\\[-6mm]
\item
They are majority starting from $1\%$ initial C's, especially in regular and single-scale networks,
and their frequencies slightly increase (especially all-C) by extending the simulations on a longer timescale
(compare regular with `l.t.' rows in the classification of Fig.~\ref{fig:r01}),
not only because of the reduced fractions of non-convergent simulations, but also because apparent nontrivial stalemates and fluctuations eventually end up in all-C or all-D.
\item
They seem to occur less frequently for $50\%$ initial C's
(see the classification of Figs.~\ref{fig:r50} and~\ref{fig:rcn}{\sffamily\textbf b},\hspace{0.2mm}{\sffamily\textbf d}),
though part of the effect is due to the slower convergence (see the fractions of non-convergent simulations and the comment below on convergence).
\item
They are more frequent at higher connectivity (compare $\langle k\rangle\hspace{-0.3mm}=4$ with $\langle k\rangle\hspace{-0.3mm}=8$) to the detriment of nontrivial stalemates and fluctuations, but also because of the faster convergence.
\item
Increasing network heterogeneity (from single-scale to scale-free networks) gives more room to nontrivial outcomes.
Moreover, all-C relatively gains over all-D with random placement of the initial C, whereas the opposite seems to occur for degree-rank-C-placement.
This is perhaps due to the higher/lower $r$-values in the first/second case.
At higher/lower game returns, D's require more/less C-neighbors to change strategy
(see condition~\eqref{eq:condrinf} in the main text),
so that convergence to all-D from $1\%$ initial C's is less/more likely.
Moreover, the evolution of cooperations is faster/slower (see below the comment below on convergence), and the fraction of non-convergent simulations is accordingly smaller/larger.
\end{itemize}

\vspace*{1mm}\noindent{\it Nontrivial stalemates}\\[-6mm]
\begin{itemize}
\item
They are infrequent and occur at low levels of C starting from $1\%$ initial C's, especially in regular and single-scale networks,
with the exception of the complete network.
And they are even less frequent if the classification is performed on a longer timescale
(compare regular with `l.t.' rows in Fig.~\ref{fig:r01}).\\[-6mm]
\item
They occur at medium-high C-levels starting from $50\%$ initial C's, and they seem to be particularly frequent in high-connected lattices
(Fig.~\ref{fig:r50}{\sffamily\textbf b})
and in low-connected scale-free networks with degree-rank-C-placement
(Fig.~\ref{fig:r50}{\sffamily\textbf i}).\\[-6mm]
\item
In general, while stalemates at low C-levels require high game returns
(to allow C's with few C-neighbors to remain C's),
stalemates at high C-levels require low returns
(not allowing the invasion of cooperation,
otherwise invasion reduces the requirement on $r$ for the further spread of C's and evolution cannot halt in a stalemate).
The latter are hence reachable only from significant initial C-levels.\\[-6mm]
\item
In the complete network, stalemates occur at the initial state
(see the mean and variance of the C-level for Fig.~\ref{fig:rcn}{\sffamily\textbf c},\hspace{0.2mm}{\sffamily\textbf d}).
Indeed C's and D's all behave in the same condition and the gap between the $r$-values above which C's do not change and D's do change is common to all individuals.\\[-6mm]
\item
The same gap is possible in high-connected lattices for game returns not allowing invasion.\\[-6mm]
\item
In scale-free networks, low returns allow a few hubs to remain defectors and the effect is even more evident under degree-rank-C-placement
(see the fraction and mean level of stalemates for Fig.~\ref{fig:r50}{\sffamily\textbf i}), because of the lower value of $r$.\\[-6mm]
\item
HK scale-free networks more easily fall in stalemate w.r.t. BA ones, at higher C-levels, essentially because HK-hubs share more common leaves and this better supports stalemates in which some D-hubs exploit leaves that nonetheless remain C's.\\[-6mm]
\item
Nontrivial stalemates are definitely less frequent at higher connectivity (compare $\langle k\rangle\hspace{-0.3mm}=4$ with $\langle k\rangle\hspace{-0.3mm}=8$), because of the increased stalemate constraints, but also because the faster convergence (see below the comment below on convergence).\\[-6mm]
\item
Contrary to what happens starting from $50\%$ initial C's, degree-rank placement reduces the possibility for nontrivial stalemates from $1\%$ initial C's (in both BA and HK scale-free networks for $\langle k\rangle\hspace{-0.3mm}=4$, compare Figs.~\ref{fig:r01}{\sffamily\textbf g},\hspace{0.2mm}{\sffamily\textbf i} and~\ref{fig:rcl}{\sffamily\textbf a},\hspace{0.2mm}{\sffamily\textbf c}), as it reduces the game return that should be sufficiently large to allow low stalemates.
\end{itemize}

\begin{figure}[t!]
\centering
\includegraphics{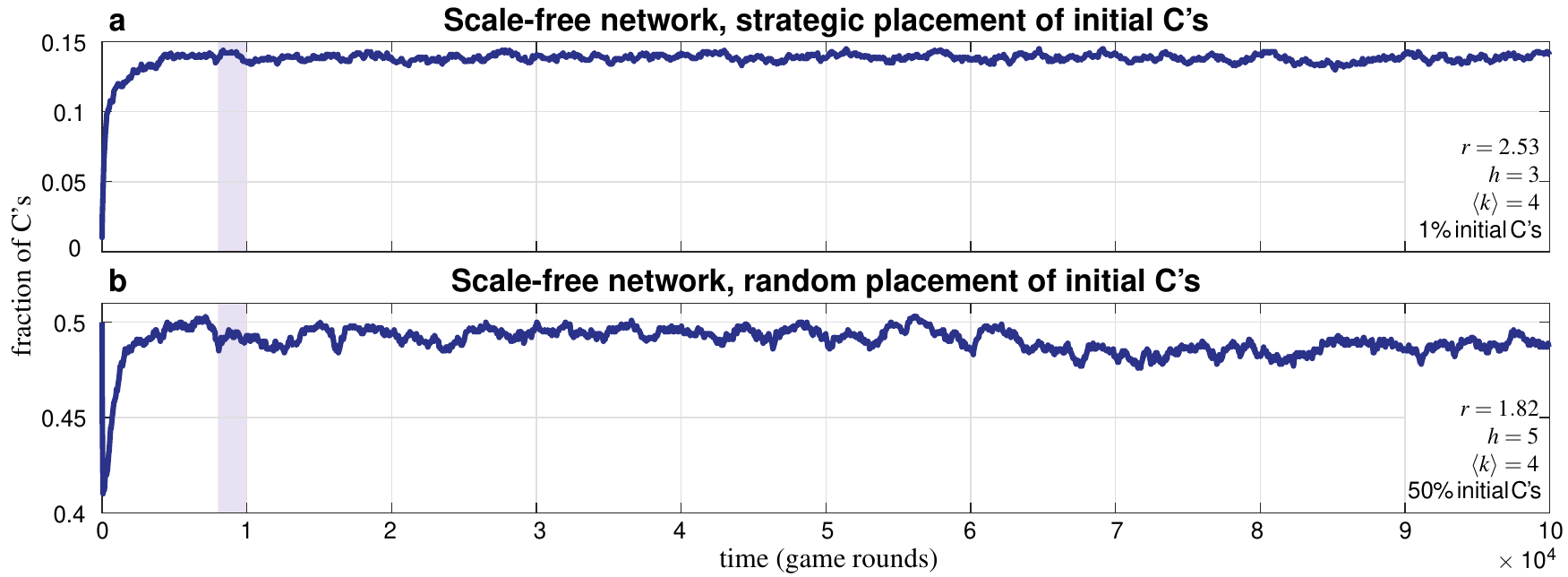}
\caption{Long-term fluctuations.
Two examples observed in scale-free (BA) networks starting from $1\%$ randomly placed ({\sffamily\textbf a}) and $50\%$ degree-rank placed ({\sffamily\textbf b}) initial C's. Other parameters: see legends and baseline values in Table~\ref{tab:par}. Both simulations are classified as long-term fluctuations in the timescale of $10^4$ game rounds, i.e., based on rounds from $0.8$ to $1$ ($\times 10^4$, shaded area); average asymptotic C-level $0.140$ ({\sffamily\textbf a}), $0.492$ ({\sffamily\textbf b}); fluctuation amplitude $0.008$ ({\sffamily\textbf a}), $0.011$ ({\sffamily\textbf b}).	
}
\label{fig:flu} 
\end{figure}

\vspace*{1mm}\noindent{\it Long-term fluctuations}\\[-6mm]
\begin{itemize}
\item
They are infrequent and occur at low/high C-levels more or less as nontrivial stalemates do.\\[-6mm]
\item
Their amplitude (the min-max excursion of the asymptotic C-level) is always very limited.
Most fluctuations are hence practical stalemates of the evolutionary process.
Figure~\ref{fig:flu} shows two relevant examples (with largest amplitude), obtained for $1\%$ ({\sffamily\textbf a}) and $50\%$ ({\sffamily\textbf b}) initial C's.\\[-6mm]
\item
Most of the comments reported for nontrivial stalemates apply to long-term fluctuations.
In particular, starting from $1\%$ initial C's, stalemates and fluctuations occur at low C-levels.
With the exception of HK scale-free networks, the C-level remains below $0.1$ on average, with peaks at $0.2$, so that the outcomes at medium-high C-levels (dots above $0.2$ in Figs.~\ref{fig:r01}, \ref{fig:rcn}{\sffamily\textbf a}, \ref{fig:red}, and \ref{fig:r2c}) most likely correspond to non-convergent simulations with increasing regression slope, that should reach all-C on a longer timescale.
This indeed happens in most of (the few) cases in Fig.~\ref{fig:r01} by ten-folding the timescale.\\[-6mm]
\item
Note that the confounded effect of the type of placement is faded if the fractions of nontrivial stalemates and fluctuations are considered together.
\end{itemize}

\vspace*{1mm}\noindent{\it Non-convergent simulations}\\[-6mm]
\begin{itemize}
\item
Starting from $1\%$ initial C's, the fractions of non-convergent simulations
(columns `n.c.$\hspace{0.1mm}+$' and `n.c.$\hspace{0.1mm}-$' in Table~\ref{tab:sim})
are sufficiently small to justify our timescale of $500$ strategy updates, on average, for each individual, with the exception of degree-$4$ ring lattices (Fig.~\ref{fig:rcn}{\sffamily\textbf a}).\\[-6mm]
\item
Our primary goal is to identify the invasion and fixation thresholds $r_{\mathrm{C},\min}^h$ and $r_{\mathrm{C},\max}^h$, i.e., starting from $1\%$ initial C's.
To allow fixation starting from low C-levels, we have therefore used a timescale a few times longer, in terms of number of individual strategy updates, than the network's diameter (see Table~\ref{tab:net}).
The degree-$4$ ring lattice is indeed the only network requiring the longer timescale.
However, looking at the dots at zero C in Fig.~\ref{fig:rcn}{\sffamily\textbf a}, we see that all-D outcomes are present up to the identified $r_{\mathrm{C},\max}^h$.\\[-6mm]
\item
Non-convergent simulations are more frequent starting from $50\%$ initial C's.
The evolution of cooperation is indeed slower, because the values of $r$ for which cooperation persists from $50\%$ are lower than those for which cooperation invades.
At low $r$, a D needs to have a significant fraction of C-neighbors to change to C (see condition~\eqref{eq:condrinf} in the main text), so she will wait longer (i.e., more strategy updates) before opting for a strategy change, compared to a case with higher $r$.
In contrast, if $r$ allows the invasion of C's, D's are willing to change strategy with few C-neighbors, and this speeds up the spread of C's.
The classification of simulations starting from $50\%$ initial C's is therefore imprecise on our timescale.
It however gives the interesting insights discussed above on nontrivial stalemates and fluctuations.\\[-6mm]
\item
The connectivity of the network speeds up the evolution of cooperation
(compare the fractions of non-convergent simulations for $\langle k\rangle\hspace{-0.3mm}=4$ and $\langle k\rangle\hspace{-0.3mm}=8$), especially if starting from $1\%$ initial C's.
More connections favor the spread of the better strategy and the higher values of $r$ required for the invasion of cooperations further contribute to the faster convergence.\\[-6mm]
\item
Network heterogeneity seems to slow down the evolution of cooperation, because of possible bottlenecks in the network structure.
The effect is amplified under degree-rank-C-placement, as the thresholds $r_{\mathrm{C},\min}^h$ and $r_{\mathrm{C},\max}^h$ are lower w.r.t. to single-scale networks.\\[-6mm]
\item
Finally, note that the majority of non-convergent simulations have positive regression slope, meaning that the C-level observed on our timescale is underestimated.
This is confirmed by the extended simulations performed for Fig.~\ref{fig:r01}
(compare regular with `l.t.' rows in Table~\ref{tab:sim}), where all-C systematically gains w.r.t. the other outcomes.
\end{itemize}

\begin{table}[t!]
\setlength\tabcolsep{1.2mm}
\centering
\scalebox{.85}{
\begin{tabular}{lllccccccc}
\hline\\[-3.5mm]
\multirow{2}{*}{\scalebox{1.16}{Fig.}} & 
\multirow{2}{*}{\scalebox{1.16}{Network}} & 
\scalebox{1.16}{Placement of} & 
\multirow{2}{*}{\scalebox{1.16}{Case}} &
\multicolumn{6}{c}{\scalebox{1.16}{Classification ($\%$)}}\\[-0.0mm]
 & & \scalebox{1.16}{initial C's} & & all-C & all-D &
stalemate  [$\langle\text{C}\rangle$, $\sigma_{\text{C}}$] \hspace*{1mm} &
fluctuation [$\langle\text{C}\rangle$,\hspace{0.1mm}$\sigma_{\text{C}}$,\hspace{0.1mm}$\langle\text{A}\rangle$,\hspace{0.1mm}$\sigma_{\text{A}}$] \hspace*{-1mm} &
n.c.$\hspace{0.1mm}+$ & n.c.$\hspace{0.1mm}-$\\[-0.0mm]
\hline\\[-3mm]
\ref{fig:r01}{\sffamily\textbf a} & planar lattice & random $1\%$ & $\langle k\rangle\hspace{-0.3mm}=4$ &
  47.50 & 51.67 & 0.39 [.004, 0] & 0 [--, --, --, --] & 0.44 & 0\\
l.t. & & & &
  47.94 & 51.67 & 0.28 [.004, 0] & 0.11 [.004, 0, .002, 0] & 0 & 0\\
\ref{fig:r01}{\sffamily\textbf b} & &                             & $\langle k\rangle\hspace{-0.3mm}=8$ &
  44.56 & 55.44 & 0 [--, --] & 0 [--, --, --, --] & 0 & 0\\
l.t. & & & &
  44.56 & 55.44 & 0 [--, --] & 0 [--, --, --, --] & 0 & 0\\
\ref{fig:r01}{\sffamily\textbf c} & single-scale    & random      & $\langle k\rangle\hspace{-0.3mm}=4$ &
  49.76 & 36.06 & 6.24 [.005, .003] & 4.53 [.010, .007, .001, .001] & 3.35 & 0.06\\
l.t. & & & &
  56.01 & 36.82 & 2.29 [.004, .003] & 4.53 [.010, .008, .003, .004] & 0.35 & 0\\
\ref{fig:r01}{\sffamily\textbf d} & &                             & $\langle k\rangle\hspace{-0.3mm}=8$ &
  55.41 & 42.97 & 0.31 [.003, .001] & 1.11 [.006, .004, .002, .001] & 0.17 & 0.03\\
l.t. & & & &
  56.28 & 43.21 & 0 [--, --] & 0.48 [.004, .002, .002, .002] & 0.03 & 0\\
\ref{fig:r01}{\sffamily\textbf e} & single-scale    & degree-rank   & $\langle k\rangle\hspace{-0.3mm}=4$ &
  60.59 & 30.23 & 3.05 [.007, .006] & 3.95 [.012, .010, .001, .001] & 2.18 & 0\\
l.t. & & & &
  64.18 & 30.95 & 0.59 [.004, .002] & 4.01 [.016, .015, .004, .005] & 0.27 & 0\\
\ref{fig:r01}{\sffamily\textbf f} & &                             & $\langle k\rangle\hspace{-0.3mm}=8$ &
  60.63 & 38.20 & 0.03 [.004, 0] & 0.97 [.005, .003, .002, .001] & 0.17 & 0\\
l.t. & & & &
  61.33 & 38.37 & 0 [--, --] & 0.30 [.007, .003, .002, .001] & 0 & 0\\
\ref{fig:r01}{\sffamily\textbf g} & scale-free      & random      & $\langle k\rangle\hspace{-0.3mm}=4$ &
  57.30 & 22.38 & 11.75 [.013, .013] & 4.78 [.019, .017, .002, .001] & 3.76 & 0.03\\
l.t. & & & &
  63.00 & 22.81 & 7.30 [.013, .014] & 6.35 [.024, .023, .002, .003] & 0.54 & 0\\
\ref{fig:r01}{\sffamily\textbf h} & &                             & $\langle k\rangle\hspace{-0.3mm}=8$ &
  63.70 & 29.53 & 1.37 [.005, .005] & 3.84 [.010, .006, .002, .001] & 1.43 & 0.13\\
l.t. & & & &
  67.23 & 30.03 & 0.54 [.004, .006] & 2.13 [.012, .007, .004, .004] & 0.07 & 0\\
\ref{fig:r01}{\sffamily\textbf i} & scale-free      & degree-rank   & $\langle k\rangle\hspace{-0.3mm}=4$ &
  30.31 & 37.00 & 5.54 [.014, .014] & 14.62 [.064, .046, .003, .002] & 10.76 & 1.77\\
l.t. & & & &
  38.69 & 37.38 & 4.08 [.012, .014] & 18.70 [.070, .047, .006, .004] & 1.15 & 0\\
\ref{fig:r01}{\sffamily\textbf j} & &                             & $\langle k\rangle\hspace{-0.3mm}=8$ &
  56.24 & 35.06 & 2.00 [.005, .002] & 4.06 [.013, .014, .002, .001] & 2.35 & 0.29\\
l.t. & & & &
  59.48 & 36.35 & 0.29 [.006, .001] & 3.76 [.016, .014, .005, .004] & 0.12 & 0\\
\hline\\[-3mm]
\ref{fig:r50}{\sffamily\textbf a} & planar lattice & random $50\%$ & $\langle k\rangle\hspace{-0.3mm}=4$ &
  12.25 & 12.50 & 3.25 [.271, .056] & 10.13 [.300, .070, .002, .001] & 38.37 & 23.50\\
\ref{fig:r50}{\sffamily\textbf b} & &                               & $\langle k\rangle\hspace{-0.3mm}=8$ &
  16.58 & 18.67 & 24.33 [.098, .090] & 3.50 [.246, .074, .002, .001] & 26.50 & 10.42\\
\ref{fig:r50}{\sffamily\textbf c} & single-scale    & random      & $\langle k\rangle\hspace{-0.3mm}=4$ &
  13.00 & 35.40 & 2.80 [.099, .148] & 14.30 [.333, .085, .003, .002] & 20.70 & 13.80\\
\ref{fig:r50}{\sffamily\textbf d} & &                             & $\langle k\rangle\hspace{-0.3mm}=8$ &
  0 & 52.00 & 1.00 [.351, 0] & 19.50 [.377, .035, .002, .001] & 7.00 & 20.50\\
\ref{fig:r50}{\sffamily\textbf e} & single-scale    & degree-rank   & $\langle k\rangle\hspace{-0.3mm}=4$ &
  64.05 & 5.72 & 4.67 [.756, .174] & 5.06 [.570, .176, .003, .002] & 12.50 & 8.00\\
\ref{fig:r50}{\sffamily\textbf f} & &                             & $\langle k\rangle\hspace{-0.3mm}=8$ &
  11.80 & 23.00 & 0.40 [.483, .127] & 13.80 [.516, .044, .002, .001] & 45.40 & 5.60\\
\ref{fig:r50}{\sffamily\textbf g} & scale-free      & random      & $\langle k\rangle\hspace{-0.3mm}=4$ &
  29.74 & 22.67 & 1.17 [.311, .280] & 6.42 [.500, .118, .004, .003] & 35.17 & 4.83\\
\ref{fig:r50}{\sffamily\textbf h} & &                             & $\langle k\rangle\hspace{-0.3mm}=8$ &
  39.88 & 17.67 & 0.56 [.427, .087] & 5.89 [.397, .072, .002, .001] & 33.11 & 2.89\\
\ref{fig:r50}{\sffamily\textbf i} & scale-free      & degree-rank   & $\langle k\rangle\hspace{-0.3mm}=4$ &
  19.60 & 10.20 & 40.10 [.861, .106] & 21.30 [.940, .057, .002, .001] & 7.00 & 1.80\\
\ref{fig:r50}{\sffamily\textbf j} & &                             & $\langle k\rangle\hspace{-0.3mm}=8$ &
  0.83 & 18.17 & 0.67 [.649, .126] & 2.33 [.719, .216, .002, .001] & 65.17 & 12.83\\
\hline\\[-3mm]
\ref{fig:rcn}{\sffamily\textbf a} & ring lattice & random $1\%$  & $\langle k\rangle\hspace{-0.3mm}=4$ &
  26.76 & 51.12 & 0 [--, --] & 0 [--, --, --, --] & 22.12 & 0\\
\ref{fig:rcn}{\sffamily\textbf b} &              & random $50\%$ & &
  10.50 & 8.83 & 9.34 [.270, .096] & 10.33 [.269, .123, .001, .001] & 55.83 & 5.17\\
\ref{fig:rcn}{\sffamily\textbf c} & complete     & random $1\%$  & &
  60.80 & 26.80 & 11.20 [.010, 0] & 1.20 [.010, .000, .002, .001] & 0 & 0\\
\ref{fig:rcn}{\sffamily\textbf d} &              & random $50\%$ & &
  0 & 0 & 99.60 [.500, .000] & 0.40 [.501, .001, .001 0] & 0 & 0\\
\hline\\[-3mm]
\ref{fig:red}{\sffamily\textbf a} & planar lattice & random $1\%$ &
\multicolumn{1}{l}{$\hspace*{-4mm}\eps=\{-1,0,.5\}$} &
  51.43 & 48.04 & 0.18 [.004 0] & 0 [--, --, --, --] & 0.35 & 0\\
\ref{fig:red}{\sffamily\textbf b} & & &
\multicolumn{1}{l}{$\hspace*{-4.3mm}\delta=\{.025,.05,.1\}$} &
  47.57 & 51.05 & 0.24 [.006 .003] & 0 [--, --, --, --] & 1.14 & 0\\
\ref{fig:red}{\sffamily\textbf c} & single-scale & random    &
\multicolumn{1}{l}{$\hspace*{-4mm}\eps=\{-1,0,.5\}$} &
  64.96 & 27.12 & 3.32 [.004, .003] & 2.80 [.009, .006, .001, .001] & 1.80 & 0\\
\ref{fig:red}{\sffamily\textbf d} & &                        &
\multicolumn{1}{l}{$\hspace*{-4.3mm}\delta=\{.025,.05,.1\}$} &
  58.35 & 31.84 & 2.81 [.004, .002] & 4.15 [.010, .008, .002, .001] & 2.85 & 0\\
\ref{fig:red}{\sffamily\textbf e} & single-scale & degree-rank &
\multicolumn{1}{l}{$\hspace*{-4mm}\eps=\{-1,0,.5\}$} &
  60.16 & 29.74 & 2.63 [.005, .005] & 3.68 [.013, .009, .002, .001] & 3.58 & 0.21\\
\ref{fig:red}{\sffamily\textbf f} & &                        &
\multicolumn{1}{l}{$\hspace*{-4.3mm}\delta=\{.025,.05,.1\}$} &
  62.21 & 27.54 & 2.79 [.005, .004] & 3.46 [.012, .009, .002, .001] & 3.92 & 0.08\\
\ref{fig:red}{\sffamily\textbf g} & scale-free & random      &
\multicolumn{1}{l}{$\hspace*{-4mm}\eps=\{-1,0,.5\}$} &
  46.80 & 30.09 & 13.37 [.012, .013] & 5.47 [.019, .018, .001, .001] & 4.27 & 0\\
\ref{fig:red}{\sffamily\textbf h} & &                        &
\multicolumn{1}{l}{$\hspace*{-4.3mm}\delta=\{.025,.05,.1\}$} &
  49.68 & 25.36 & 14.89 [.011, .011] & 5.21 [.019, .019, .001, .001] & 4.86 & 0\\
\ref{fig:red}{\sffamily\textbf i} & scale-free & degree-rank   &
\multicolumn{1}{l}{$\hspace*{-4mm}\eps=\{-1,0,.5\}$} &
  32.93 & 33.36 & 6.14 [.020, .024] & 14.71 [.057, .038, .002, .001] & 11.50 & 1.36\\
\ref{fig:red}{\sffamily\textbf j} & &                        &
\multicolumn{1}{l}{$\hspace*{-4.3mm}\delta=\{.025,.05,.1\}$} &
  45.22 & 27.67 & 5.61 [.019, .027] & 12.06 [.054, .038, .002, .001] & 8.72 & 0.72\\
\hline\\[-3mm]
\ref{fig:r2c}{\sffamily\textbf a} & planar lattice & random pair $1\%$\hspace*{-6mm} & $\langle k\rangle\hspace{-0.3mm}=4$ &
  20.14 & 73.64 & 2.29 [.006, .002] & 0.64 [.012, .008, .002, .001] & 3.29 & 0\\
\ref{fig:r2c}{\sffamily\textbf b} & &                             & $\langle k\rangle\hspace{-0.3mm}=8$ &
  46.69 & 53.25 & 0.03 [.004, 0] & 0 [--, --, --, --] & 0.03 & 0\\
\ref{fig:r2c}{\sffamily\textbf c} & single-scale    & random pair & $\langle k\rangle\hspace{-0.3mm}=4$ &
  27.42 & 47.50 & 8.83 [.006, .004] & 9.92 [.012, .010, .002, .001] & 6.04 & 0.29\\
\ref{fig:r2c}{\sffamily\textbf d} & &                             & $\langle k\rangle\hspace{-0.3mm}=8$ &
  53.59 & 42.56 & 0.49 [.004, .002] & 2.48 [.008, .007, .003, .002] & 0.81 & 0.07\\
\ref{fig:r2c}{\sffamily\textbf e} & scale-free      & random pair & $\langle k\rangle\hspace{-0.3mm}=4$ &
  37.26 & 27.06 & 15.55 [.017, .019] & 12.24 [.030, .029, .001, .001] & 7.24 & 0.65\\
\ref{fig:r2c}{\sffamily\textbf f} & &                             & $\langle k\rangle\hspace{-0.3mm}=8$ &
  42.48 & 38.12 & 2.08 [.005, .002] & 11.60 [.012, .009, .002, .001] & 5.32 & 0.40\\
\hline\\[-3mm]
\ref{fig:rcl}{\sffamily\textbf a} & HK scale-free   & random $1\%$\hspace*{-6mm} & $\langle k\rangle\hspace{-0.3mm}=4$ &
  35.90 & 13.82 & 45.71 [.133, .230] & 0.85 [.089, .164, .002, .001] & 3.64 & 0.08\\
\ref{fig:rcl}{\sffamily\textbf b} & &                             & $\langle k\rangle\hspace{-0.3mm}=8$ &
  53.17 & 42.63 & 1.14 [.004, .002] & 2.40 [.009, .006, .002, .001] & 0.66 & 0\\
\ref{fig:rcl}{\sffamily\textbf c} & HK scale-free   & degree-rank   & $\langle k\rangle\hspace{-0.3mm}=4$ &
  41.23 & 10.97 & 34.69 [.524, .375] & 4.80 [.270, .235, .002, .002] & 8.14 & 0.17\\
\ref{fig:rcl}{\sffamily\textbf d} & &                             & $\langle k\rangle\hspace{-0.3mm}=8$ &
  32.04 & 55.81 & 3.77 [.005, .003] & 5.92 [.013, .017, .002, .001] & 2.15 & 0.31\\
\hline
\end{tabular}}
\caption{Classifications of the simulations performed for $r_{\mathrm{C},\min}^h<r<r_{\mathrm{C},\max}^h$ grouped by panel of Figs.~\ref{fig:r01} and~\ref{fig:r50}--\ref{fig:rcl}.
For each panel, the table reports the fractions of the six possible outcomes (all-C, all-D, stalemate, fluctuation, non-convergence with positive/negative regression slope).
To quantify stalemates and fluctuations, the table reports the mean and standard deviation ($\langle\text{C}\rangle$ and $\sigma_{\text{C}}$)
of the average asymptotic C-level and, only for fluctuations, the mean and standard deviation ($\langle\text{A}\rangle$ and $\sigma_{\text{A}}$)
of the oscillation's amplitude (the min-max excursion of the asymptotic C-level).
To validate the classification, the simulations of Fig.~\ref{fig:r01} have been extended over a ten-times longer timescale
($5000/\delta=10^5$ game rounds, the last $20\%$ of which used for the classification; see row label `l.t.'), showing no significant change.}
\label{tab:sim}
\end{table}

\clearpage

\bibliographystyleSI{naturemag}
\bibliographySI{egtnetbib}

\end{document}